\documentclass{article}
\usepackage{amsfonts,amsmath,amssymb}
\usepackage{mathrsfs}
\usepackage{nicefrac}
\usepackage{graphicx}
\usepackage{color}
\usepackage[compress]{cite}
\usepackage{hyperref}
\usepackage{accents}
\usepackage{varioref}
\usepackage{subfigure}
\usepackage[normalem]{ulem}
\usepackage{hyperref}
\hypersetup{
    colorlinks=true, 
    linktoc=page,    
    linkcolor=blue,  
    urlcolor=magenta,
    citecolor=blue
}
\newcommand{\be}{\nopagebreak[3]\begin{equation}}
\newcommand{\ee}{\end{equation}}
\newcommand{\ba}{\nopagebreak[3]\begin{eqnarray}}
\newcommand{\ea}{\end{eqnarray}}
\newcommand{\la}{\label}

\labelformat{section}{Section #1} 
\labelformat{subsection}{Section #1} 
\labelformat{subsubsection}{Section #1}
\labelformat{subsubsubsection}{Section #1}
\labelformat{equation}{Eq.~(#1)} 
\labelformat{figure}{Fig.~#1} 
\labelformat{subfigure}{Fig.~\thefigure#1} 
\labelformat{table}{Table.~#1} 
\labelformat{appendix}{Appendix #1}

\title{\bf Echoes from the braneworld black holes}

\author{Ramit Dey\footnote{ramitdey@gmail.com}$~^{1,2}$, Sumanta Chakraborty\footnote{sumantac.physics@gmail.com}$~^{1,3}$ and 
Niayesh Afshordi\footnote{nafshordi@pitp.ca }$~^{2,4,5}$
\\
{$^{1}$\small{School of Physical Sciences, Indian Association for the Cultivation of Science, Kolkata-700032, India}}
\\
{$^{2}$\small{Perimeter Institute For Theoretical Physics, 31 Caroline St N, Waterloo, Canada}}
\\
{$^{3}$\small{School of Mathematical and Computational Sciences}}
\\
{\small{Indian Association for the Cultivation of Science, Kolkata-700032, India}}
\\
{$^{4}$\small{Department of Physics and Astronomy}}
\\
{\small{University of Waterloo, 200 University Ave W, N2L 3G1, Waterloo, Canada}}
\\
{$^{5}$\small{Waterloo Centre for Astrophysics, University of Waterloo, Waterloo, ON, N2L 3G1, Canada}}}
\date{}
\begin{document}

\maketitle

\begin{abstract}
The holographic interpretation of the Randall-Sundrum model based on the adaptation of the AdS/CFT correspondence to the braneworld scenario states that the black holes localized on the brane are {\it quantum corrected}. This can be better understood from the fact that the classical $AdS_{5}$ bulk dynamics is dual to gravity coupled with CFT on the four dimensional brane. Based on the back reaction of the CFT on the classical black hole geometry, localized on the brane, it is expected that there exist possible near horizon modifications. This may result in the black hole horizon becoming partially reflective, thus giving rise to echoes in the ringdown signal in gravitational wave observations. In this paper, we investigate the existence of such echoes in the ringdown phase of a black hole localized on the brane, carrying a \emph{negative} tidal charge, and establish the layout for future investigation of higher dimensional effects in the ringdown signal. Confirmed detections of echoes at the current levels of instrumental sensitivity can constrain the dimensionless value of tidal charge to $|Q| \lesssim M^2$ . 
\end{abstract}

\section{Introduction}

The detection of gravitational waves \cite{Abbott:2016blz,TheLIGOScientific:2016pea,Abbott:2016nmj,TheLIGOScientific:2016src,Abbott:2017vtc} due to the coalescence of compact binary objects opened the door to test various non-trivial aspects of spacetime geometry in the strong gravity regime. Improved and/or next generation of detectors, will make it possible to probe deeper inside the signal from the merger event, giving us an insight about the strong gravity regime dominated by possible quantum effects, or novel semi-classical physics.

Out of the three different phases of a binary merger, based on the completely different physics behind each one of them, the ringdown phase is of particular interest to us. Broadly speaking, the ringdown phase involves the final relaxation of the composite object, formed due to collision of two binary objects, in this case black holes, as it settles down to a final equilibrium configuration. The ringdown phase is expected to shed light on the nature of the event horizon and can help us either verify the classical {\it no-hair theorem}, or rather discover non-trivial quantum effects at play close to the horizon \cite{PhysRevLett.116.171101}. 

In classical general relativity, the final ringdown phase is primarily dominated by the quasi normal modes (QNMs) of the black hole (e.g., \cite{Kokkotas:1999bd}). For obtaining the QNMs one generally imposes the following boundary conditions: purely outgoing waves at spatial infinity and purely ingoing waves at the horizon. The situation may change drastically if quantum effects near the horizon are taken into account. In particular, due to various quantum effects near the horizon (e.g.,  fuzzball \cite{Mathur:2005zp}, firewalls \cite{Almheiri:2012rt} scenario, Kerr wormholes \cite{Bueno:2017hyj}) or upon interpreting the black hole as a multilevel quantum system \cite{Oshita:2019sat}, the ingoing waves can get partially reflected from a region very close to the horizon and reach back to the asymptotic observer at infinity with a time delay. This produces an echo like pattern in the spectrum of the gravitational waves, where the echo signals are separated from the primary ringdown signal by the associated time delay due to reflection \cite{PhysRevLett.116.171101,PhysRevD.94.084031,Abedi:2016hgu,Oshita:2018fqu,Wang:2019rcf}. Several studies claimed to have found potential evidence of echoes in the LIGO data \cite{Abedi:2016hgu,Wang_2018,Conklin:2017lwb,abedi2018echoes,Conklin:2019fcs,Holdom:2019bdv}, although such evidences remains controversial (e.g., \cite{Westerweck:2017hus,Tsang:2019zra,Salemi:2019uea,Uchikata:2019frs}).  

Given the significant modifications made to the ringdown phase, pertaining to the reflective boundary condition at the horizon, it is instructive to understand whether there exists any model where such reflective boundary conditions are a necessity. We demonstrate in this work that such is indeed the scenario when one introduces extra spatial dimensions to the spacetime. In particular, the reflective boundary conditions become essential when brane localized black holes in the Randall-Sundrum braneworld paradigm are considered. As in the standard scenario with an extra spatial dimension, the matter fields are localized on a four dimensional brane and only gravity can propagate in the five dimensional bulk spacetime \cite{Randall:1999ee,Randall:1999vf,Maartens:2010ar}. Such an approach calls for obtaining the effective gravitational field equations on the four dimensional brane, which was derived using the Gauss-Codazzi formalism in \cite{Shiromizu:1999wj,Harko:2004ui,Aliev:2005bi,Chakraborty:2015taq,Chakraborty:2014xla}. The effective gravitational field equations so obtained, can be solved for a static and spherically symmetric spacetime, yielding a metric which resembles a Reissner-Nordstr\"{o}m black hole with a `tidal charge' instead of the usual electric charge \cite{Dadhich:2000am}. Due to non-trivial effects inherited from the bulk, the effective gravitational field equations on the brane are modified and these modifications are manifested in the brane localized black hole solution as the tidal charge. This motivates one to look if the presence of the tidal charge due to higher spacetime dimensions can have any observational significance. Several analysis has been performed in various observational avenues to look for the presence of this negative tidal charge \cite{Banerjee:2019nnj,Banerjee:2019sae,Banerjee:2017hzw,Vagnozzi:2019apd}, including its effect on gravitational waves \cite{Chakravarti:2019aup,Chakravarti:2018vlt,Chakraborty:2017qve,Visinelli:2017bny,Rahman:2018oso,Mishra:2019ged}. However, it should be emphasized that it is difficult to obtain static black hole solutions that are localized on the 3-brane and free of any pathology when extended to the bulk \cite{Chamblin_2000,Bruni_2001,Dadhich:2000am,Chamblin_2001,Kanti_2002,Emparan_2000,Gregory}.

From the adaptation of the AdS/CFT conjecture to the braneworld model \cite{Randall:1999vf} it was argued that solving the classical five dimensional Einstein's equations in the bulk are equivalent to solving the semi-classical Einstein's equations in the four dimensional dual CFT theory coupled with gravity on the brane \cite{Emparan:2002px,Anderson_2005,Fabbri_2007}. Therefore, on the brane hypersurface, the CFT stress energy tensor would induce quantum effects, hence affecting the black hole solutions localized on the brane. Thus one may conclude that the black hole solutions derived from the classical bulk gravitational field equations will inherit non-trivial quantum corrections on the brane. 

Based on the above discussion, we will consider the brane localized black hole solution presented in \cite{Dadhich:2000am} as a quantum black hole and compute the QNMs due to scalar perturbation by imposing reflective boundary condition at the horizon, rather than purely ingoing boundary condition. In particular, we solve for the propagation of the massless scalar field modes, whose evolution is governed by the Klein-Gordan equation in the above background geometry of a braneworld black hole. As we will demonstrate, the QNMs derived in such an approach exhibit echoes. 
 
The paper is organized as follows: In \ref{Sec_2} we give the general form of the scalar perturbation equation in a static and spherically symmetric spacetime that we will solve analytically, with a low frequency approximation, as well as numerically in order to determine the ringdown modes and hence the nature of echoes. Subsequently, in \ref{Sec_3} we have reviewed the static, spherically symmetric black hole solution localized on a 3-brane and hence  motivate the necessity of reflective boundary condition at the horizon. The real and the imaginary parts of the frequencies associated with QNMs using an analytic method have been presented in \ref{Sec_4}. Finally, in \ref{Sec_5} we have performed a  numerical analysis in order to obtain the frequencies of the QNMs and show how the quantum black holes in the braneworld scenario would give rise to echoes in the ringdown. We further comment on the dependence of the ringdown spectrum on the tidal charge of the black hole as well as on future directions of exploration. 

\section{Scalar field in a static and spherically symmetric spacetime}\label{Sec_2}

Our objective in this paper is to study the scalar perturbation of a black hole localized on the brane, which presumably experiences quantum corrections near the horizon, resulting in partial refelction of the in-falling modes which appear as echoes to the asymptotic observer at infinity. To arrive at the main conclusion, we will keep our analysis simple and consider perturbations due to a massless scalar field living in this background geometry of a brane localized black hole. This is  because the qualitative predictions for higher spin perturbations would be more or less identical with the much simpler scalar perturbation and in this paper, our primary objective is to motivate the scenario in which one can justify the existence of  the quantum black holes that would give rise to such echoes in the ringdown spectrum.

Given this preamble, in this section, we will briefly discuss the basic ingredients necessary to compute how a scalar field evolves in a static and spherically symmetric spacetime. The results derived in this section will be useful in the later parts of this work when we apply the formalism to a specific solution of the effective gravitational field equations on the brane. The scalar field will be assumed to be a free field, i.e., without any potential term and without any mass. Thus the evolution of the scalar field is determined by the Klein-Gordon equation in the static background spacetime, respecting spherical symmetry. As we will see, the Klein-Gordon equation reduces to a linear second order differential equation on which appropriate boundary conditions must be imposed. Then one finds that the evolution of the scalar field must exhibit certain characteristic time scales, whose inverses are known as the QNMs associated with the spacetime. 

For generality, we start by writing down the generic metric ansatz depicting a static and spherically symmetric spacetime, which takes the following form, 
\ba\label{eq1}
dS^2=-f(r)dt^2+\frac{dr^2}{h(r)}+r^2(d\theta^2+sin^2\theta d\phi^2)~.
\ea
Here, $f(r)$ and $h(r)$ are arbitrary functions of the radial coordinate. In this background geometry, the propagation of the scalar field $\psi$ is governed by the equation $\square \psi=0$. Due to spherical symmetry and staticity of the spacetime, it is possible to decompose the scalar field $\Psi$ as
\ba
\Psi(t,r,\theta,\phi)=\frac{1}{r}\sum_{\ell=0}^{\infty} \sum_{m=-\ell}^{\ell} 
e^{-i\omega t}Y_{\ell m}(\theta,\phi)\psi_{\ell m}(r)~,
\ea
where $Y_{\ell m}(\theta,\phi)$ corresponds to the spherical harmonics. Inserting the above expansion of the scalar field in its field equation, i.e., $\square \psi=0$, we obtain
\ba \la{wave_eq}
\sqrt{f(r)h(r)}~\partial_r\left(\sqrt{f(r)h(r)}~\partial_r \psi_{\ell m}\right)
+\left\{\omega^2-V_{\ell}(r)\right\}\psi_{\ell m}=0~,
\ea
where the potential $V_{\ell}(r)$ appearing in the above expression is given as 
\ba \la{potential}
V_{\ell}(r)=f(r)\bigg[ \frac{\ell(\ell+1)}{r^2}+\frac{1}{2r}\bigg(h'(r)+\frac{h(r)f'(r)}{f(r)}\bigg)\bigg]~.
\ea
Thus the Klein-Gordon equation can be reduced to an ordinary differential equation for $\psi_{\ell m}(r)$ in the radial coordinates. It is advantageous to re-scale the field $\psi_{\ell m}(r)$, such that $\psi_{\ell m(r)}=rR_{\ell m}(r)$. Thus the differential equation satisfied by the function $R_{\ell m}(r)$ immediately follows from \ref{wave_eq} as,
\ba\la{wave_eq_mod}
\sqrt{f(r)h(r)}\partial_r \bigg(r^2\sqrt{f(r)h(r)}\partial_r R_{\ell m}\bigg)
+\bigg[r^2 \omega^2-f(r)\ell(\ell+1) \bigg]R_{\ell m}=0.
\ea
Thus given any spacetime geometry one can express the above equation as a second order differential equation in the radial coordinate. In order to determine the solution it is necessary to impose suitable boundary conditions which tells us how the field asymptotes. Using such a set of appropriate boundary conditions, from the poles of the Green's function of the above equation one can determine the frequencies associated with the QNMs of the system. This is the approach we will take in the later sections, where the above differential equation will be used to determine the associated QNMs with appropriate boundary conditions.  

\section{Vacuum solution on the brane}\label{Sec_3}

It is difficult to obtain static black hole solutions on the brane and the first attempt was made in \cite{Chamblin_2000} by replacing the Minkowski metric on the brane by a Schwarzschild metric. Even though one would expect the black hole horizon to extend into the bulk, it is desirable that the singularity should always stay confined on the brane. The problem with the black string solution of \cite{Chamblin_2000} is that it extends all the way up to the AdS horizon. Also, it was shown that the solution suffers from classical instability for a certain range of parameter space \cite{Gregory_1993, Gregory_2000}. Due to such complications arising out of extension of black hole horizon in the bulk and possible instability of the spacetime, an exact solution for a black hole localized on a four dimensional brane is not known and it has also been proposed that a static classical black hole formed by gravitational collapse does not exist on the brane \cite{Bruni_2001}. For this reason, various approximate methods are used to gain a better understanding of the black holes localized on the brane. In this section, we will describe one such way to arrive at vacuum solutions that are localized on the brane. This corresponds to finding out the `effective' gravitational field equations from the perspective of a brane observer and then determining its solution.

In order to derive such effective gravitational field equations, one starts with the gravitational field equations in the bulk, which is taken to be the five dimensional Einstein's equations with a negative bulk cosmological constant $\Lambda$. The gravitational field equations on the four-dimensional brane are obtained by projecting various geometrical quantities, appearing in the bulk gravitational field equations, onto the brane. This is achieved by using the projector $h^{A}_{B}=\delta^{A}_{B}-n^{A}n_{B}$, where $n_{A}$ is the unit normal, perpendicular to the four-dimensional hypersurface, i.e., normal to the brane. The fact that $h^{A}_{B}$ is a projector can be understood by noting that it satisfies the following properties: $n_{A}h^{A}_{B}=0$ and $h^{A}_{C}h^{C}_{B}=h^{A}_{B}$. Thus one uses this projector to define various geometrical quantities intrinsic to the brane hypersurface. In the context of curvature tensor, such a relation between intrinsic curvature on the brane with bulk curvature is given by the Gauss-Codazzi equations. This can be further contracted to determine the projection of the bulk Einstein's equations on the brane. Since we are considering the case of vacuum brane, there is no matter energy momentum tensor on the brane. However, there is a non-zero brane tension $\lambda_{\rm b}=\sqrt{-(6\Lambda/8\pi G_{5})}$, where $G_{5}$ is the five dimensional gravitational constant. This choice of $\lambda_{b}$ ensures that the `effective' four dimensional cosmological constant identically vanishes. Thus the `effective' gravitational field equations on the vacuum brane take the following form,
\ba  \la{equation}
~^{(4)}G_{ab}+E_{ab}=0~.
\ea
In the above expression ${}^{(4)}G_{ab}$ is the Einstein tensor constructed exclusively using the metric induced on the brane, i.e., $h_{ab}$. The other quantity appearing in \ref{equation} corresponds to $E_{ab}\equiv W_{ACBD}e^{A}_{a}n^{C}e^{B}_{b}n^{D}$, where $W_{ABCD}$ is the bulk Weyl tensor and $e^{A}_{a}\equiv (\partial x^{A}/\partial y^{a})$ is another way of expressing the projector. Even though the bulk gravitational field equations were purely Einstein in nature, the effective theory on the brane inherits additional corrections over and above the Einstein term and is related to the Weyl tensor in the bulk. 

\subsection{Exact solution with negative tidal charge}

It turns out that it is indeed possible to solve \ref{equation} in the context of static and spherically symmetric vacuum four dimensional spacetime. The metric derived with the above symmetry requirements has the following line element \cite{Dadhich:2000am}
\ba  \la{metric}
ds^2=-\left(1-\frac{2M}{r}-\frac{Q}{r^2}\right)dt^2+\left(1-\frac{2M}{r}-\frac{Q}{r^2}\right)^{-1}dr^{2}
+r^2\left(d\theta^2+\sin^2\theta d\phi^2\right)~.
\ea
The term $(Q/r^{2})$ is originating from the projection of the bulk Weyl tensor, i.e., $E_{ab}$, the correction term in the effective gravitational field equations. The metric structure resembles the Reissner-Nordstr\"{o}m solution, since the tensor $E_{ab}$ coming from bulk geometry has a structure, similar to that of the energy momentum tensor of a Maxwell field. Thus the charge parameter $Q$ must have its origin in the bulk spacetime, whose understanding requires extending the brane solution into the bulk. This can be achieved by starting with an ansatz for the bulk spacetime, which can reproduce the correct solution on the brane and then solving for the unknown functions using the evolution equation for the extrinsic curvature. Such an evolution is not possible analytically, rather one must look for numerical techniques to solve them. This was achieved in \cite{Chamblin_2001}, where care was taken so that at each stage of the evolution the Hamiltonian and momentum constraint equations were satisfied. Such an analysis reveals that the horizon indeed penetrates into the bulk upto certain distance, which depends on the tidal charge parameter $Q$. In particular, as the tidal charge parameter increases, the extent of the horizon in the bulk spacetime decreases, i.e., the black hole becomes more and more localized. Thus the tidal charge $Q$ can be taken to be a parameter estimating extension of the black hole horizon in the bulk spacetime. As the numerical analysis of \cite{Chamblin_2001} demonstrates, as far as the evolution of the brane into the bulk is considered, there is no restriction on the tidal charge $Q$ from bulk dynamics.

\begin{figure}[h]
\centering
\includegraphics[width=12cm]{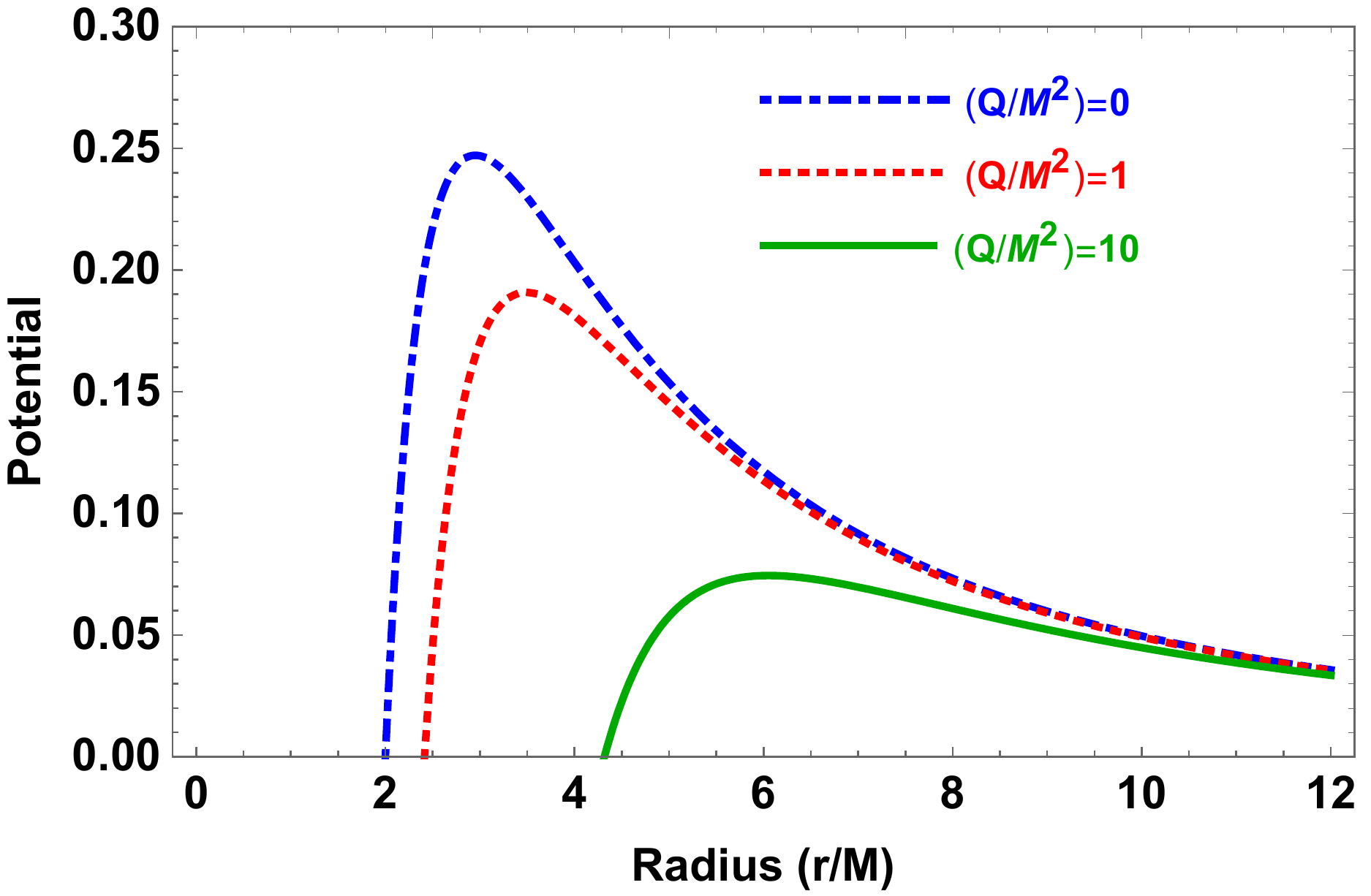}
\caption{The maxima of the potential is at $r_{\rm max}\sim 3M$, for $Q=0$, but as we can see, with increasing value of the tidal charge, the peak of the potential decreases and the maxima shifts towards larger and larger values of the radial coordinate. This will have serious implications for the echoes.}\label{fig:potential}
\end{figure}

It should be emphasized that, despite the superficial similarity of the metric presented in \ref{metric} with the Reissner-Nordstr\"{o}m solution, the solution is actually associated with vacuum spacetime, where $Q$ is appearing (with a distinctive \emph{negative} sign before it) due to existence of extra dimension. To reiterate, in Reissner-Nordstr\"{o}m metric the coefficient of $(1/r^{2})$ term, defined as the electric charge, is always positive, while here the coefficient of $(1/r^{2})$ term can have negative sign and it is defined as the `tidal' charge \footnote{It enters the metric as an integration constant but one can think of the tidal charge as an effective charge parameter containing the information about the extra dimension.}. In particular when $Q>0$, the coefficient of $(1/r^{2})$ term has a negative sign and is a distinctive signature of the presence of higher spacetime dimensions. Given the structure of the metric as depicted in \ref{metric}, one can easily read off, $f(r)=h(r)=1-(2M/r)-(Q/r^{2})$ and use this functional form in \ref{potential} in order to determine the potential, a massless scalar field experiences when living on this black hole spacetime. For illustrative purposes and keeping in mind future applications we have depicted the potential associated with the evolution of a massless scalar field in the above black hole background for different values of the tidal charge $Q$ in \ref{fig:potential}. As evident from \ref{fig:potential}, with larger and larger values of the tidal charge parameter $Q$, the height of the potential decreases and the location of the maxima of the potential shifts to the larger radial distances. As we will demonstrate, these features will have significant implications on the ringdown phase of the black hole.  

From the perspective of an observer located on the brane hypersurface, the above metric with positive $Q$, i.e., negative tidal charge inherits a single horizon located at $r_{\rm h}=M+\sqrt{M^2+Q}$. Thus one may interpret this solution as a black hole solution with a horizon, from the perspective of a four dimensional observer. The above result also demonstrates another non-trivial feature associated with this solution as it admits a single horizon, unlike the Reissner-Nordstr\"{o}m solution, which possesses two horizons. Further note that unlike the case of Reissner-Nordstr\"{o}m black hole, the existence of horizon does not enforce any condition on the tidal charge parameter $Q$. The only constraint on $Q$ may come from local physics, e.g., perihelion precession and bending angle measurement from sun, where one obtains, $(Q/M^{2})\lesssim 0.1$ \cite{Bhattacharya:2016naa}. However, this assumes a certain behaviour for the interior of the star, which need not be true. It is even possible to design the stellar interior, such that the external spacetime is identical to Schwarzschild. However, for black holes the external geometry is still given by \ref{metric}. Hence it is indeed possible that black holes have larger values of the tidal charge parameter. Further hint for larger value for tidal charge parameter follows from black hole shadow measurement, where the observations were consistent with $Q/M^{2}\sim 1$ \cite{Banerjee:2019nnj}. Thus a priori larger values for the tidal charge parameter cannot be ruled out. Further, in the AdS/CFT perspective such a tidal charge parameter must have its origin in the backreaction of the CFT on the brane, which makes $Q$ related to energy scale of the CFT with the possibility of running in the strong gravity regime. This also makes a strong case for larger values of the tidal charge parameter $Q$.

In order to infer the true nature of this horizon, i.e., whether it is an event horizon, one needs the global structure of the spacetime, which includes the bulk geometry as well. In the context of the brane metric presented in \ref{metric} one can integrate the evolution equations for the extrinsic curvatures into the bulk, while the constraint equations take care of the accuracy of the method. Since the bulk inherits a negative cosmological constant the volume decreases exponentially as one probe more and more into the bulk geometry, thus after certain length scale into the bulk the numerical evolution cannot be continued any further. However, it turns out that in the above mentioned scheme of evolution into bulk, the surface $r=r_{\rm h}$ appears as an apparent horizon, i.e., the outermost surface having a negative expansion for the outgoing null geodesic congruence \cite{Chamblin_2001}. 

\subsection{Quantum corrections at the horizon}

As we have demonstrated earlier, the extension of the black hole spacetime on the brane hypersurface, presented in \ref{metric}, into the bulk geometry makes the surface $r=r_{\rm h}$ an apparent horizon. Thus it is no longer justified to introduce a purely ingoing boundary condition at $r=r_{\rm h}$. Further motivation and support for the above statement originates from considering the AdS/CFT duality, according to which the boundary theory of an AdS bulk is a CFT. In the RS braneworld scenario, we have a similar setup where our universe is assumed to be a hypersurface in a $AdS_{5}$ bulk spacetime. As a consequence of this, the zero mode of the five dimensional gravity gets trapped on the brane, inducing four dimensional gravity coupled to the dual CFT on the brane. Thus the holographic interpretation of the RS scenario states that the classical bulk dynamics is dual to four dimensional gravity coupled to a cut-off CFT on the brane. This would imply that black holes localized on the brane are always quantum corrected, since there would be a back reaction on them due to the CFT coupling \cite{Emparan:2002px,Anderson_2005,Fabbri_2007}. The quantum interpretation of the black holes on the brane further explains why static solutions were hard to obtain on the brane which has a pathology free bulk extension. It was further shown that the correction to Newtonian potential in the RS scenario \cite{Garriga_2000} can be accounted from the one loop correction to the graviton propagator \cite{Duff:2000mt}. 

Following such discussions regarding quantum corrections at the horizon for brane localized black hole, in \cite{Fabbri_2007} a plausible `quantum' modification of the horizon, due to the bulk dynamics, was studied. This was achieved by investigating the null geodesics related to the massive KK mode trajectories in the bulk. It was demonstrated that the apparent horizon of the brane localized black hole indeed gets shifted due to quantum corrections and is given as
\ba \la{difference}
\Delta r_{\rm h}^{\rm (brane)}\equiv \epsilon \sim \frac{N^2l_{\rm p}^2}{M}~,
\ea
where $l_{\rm p}$ is the four dimensional Planck length, $M$ is the black hole mass and $N$ are the degrees of freedom of the dual CFT living on the brane. Note that the above result is for $G=1=c$ unit. If we want to restore the units, then, the black hole mass must be replaced by $GM/c^{2}$. For a black hole of mass $1 M_{\odot}$, the quantity $GM/c^{2}=1.48\times 10^{3}~\textrm{m}$, while $GM/c^{3}=0.49\times 10^{-5}~\textrm{s}$. Using the holographic principle, the effective CFT degrees of freedom $N$ can be written as
\ba
N^2 \sim \left(\frac{L}{l_{\rm p}}\right)^2\sim 10^{30}\left(\frac{L}{1~\textrm{mm}}\right)^{2}~,
\ea 
where $L$ is the bulk AdS radius and one can impose bounds on $L$ from various observation since it would also determine the lifetime of a black hole \cite{Emparan_2003} as well as the extent of the extra dimension. Other evidence for the quantum corrected black holes based on the holographic interpretation of RS2 was provided in \cite{Anderson_2005}, where it was argued that the quantum corrections due to back-reaction will act on the Schwarzschild metric localized on the brane and thus would modify the classical horizon. 

\subsection{Boundary condition at the horizon}

Having elaborated on the black hole solution derived from the effective gravitational field equations as well as possible quantum modifications to the horizon we now turn our attention to the discussion of the boundary condition at the apparent horizon and how it may depend on the quantum effects near the horizon. The scalar field as well as the horizon is affected by the existence of bulk spacetime, through either the potential in the Klein-Gordon equation or quantum corrections. Thus it is safe to assume that the boundary condition at the (apparent) horizon need not be strictly ingoing as it would have been for the case for an event horizon in a  static black hole spacetime. 

To quantify this speculation, we assume that the black hole will partially reflect the ingoing scalar modes. Further to make things simple, we will assume following \cite{Abedi:2016hgu,Oshita:2018fqu,Wang:2019rcf} and the discussion leading to \ref{difference}, that there is a membrane outside the horizon, at $r_0\sim r_{\rm h}+\epsilon$ due to some quantum effects, possibly originating from the back-reaction of the CFT on the brane, a consequence of the AdS/CFT correspondence. Such quantum effects, naturally leads to reflecting boundary conditions at $r_{0}$ (at least for frequencies $\lesssim$ Hawking temperature \cite{Oshita:2018fqu}). To be consistent with the semi-classical gravity on the brane, we place the membrane at $\eta$ times the Planck proper length away from the horizon, motivated by \ref{difference}. Thus we have the following relation determining the parameter $\eta$, which encapsulates the modifications to the location of the horizon, as,
\ba
\int^{r_{\rm h}+\epsilon}_{r_{\rm h}}\sqrt{g_{rr}}dr\sim \eta l_{\rm p}~.
\ea
Solving the integral in the near horizon limit, one can fix the location of the membrane\footnote{Even though the black string solution used in \cite{Fabbri_2007}, to conclude about the quantum ergosphere, does not take into consideration the effect of the tidal charge, we use the induced black hole metric on the brane with the tidal charge parameter to make the argument more general while capturing the quantitative behaviour of the presence of any quantum effects in the near horizon region. Also, in $Q\sim0$ limit the black hole metric, as presented in \ref{metric} reduces to the form that was used in \cite{Chamblin_2000,Fabbri_2007}.}, i.e., $\eta$ in terms of black hole parameters and Planck length $l_{\rm p}$ as
\ba
\eta^{2}\sim \frac{N^{2}}{M}\frac{4r_{\rm h}^2}{(r_{\rm h}-r')}=2N^{2}\frac{\left(1+\sqrt{1+(Q/M^{2})}\right)^{2}}{\sqrt{1+(Q/M^{2})}}~,
\ea
where $r'\equiv M-\sqrt{M^2+Q}$ is introduced in the above expression for later convenience. Even though $r'$ has similarity with the inner horizon of a Reissner-Nordstr\"{o}m black hole, in the present context with negative tidal charge, i.e., positive values of $Q$, it does not represent any physical length scale associated with the system. It further follows from the above expression that $\eta^{2}\sim N^{2}$ and for $L=1\textrm{mm}$, it follows that $\eta^{2}\sim 10^{30}$. Thus the membrane is located at $\sim 10^{-18}~\textrm{m}$. Note that the smaller the AdS radius, the smaller the value of $\eta$ and hence the location of the membrane depends strongly on the AdS curvature scale. Thus the presence of extra dimension and its length indeed affects the location of the membrane and as we will demonstrate, this in turn will affect the echo spectrum. Hence echoes have a direct correspondence with the length of the extra dimension.

\section{Structure of quasi-normal modes and echo}\label{Sec_4}

In this section, we solve for the QNMs associated with the scalar field perturbation in the background spacetime described by the metric presented in \ref{metric}. In the case of classical black holes the boundary condition at the horizon will be ingoing (for such a boundary condition the QNM is calculated in \cite{Toshmatov:2016bsb}), however as argued above the apparent horizon $r=r_{\rm h}$ will be modified due to quantum effects from the coupled CFT, so it gives us the freedom to introduce a reflecting boundary condition at the horizon, or at a surface Planck proper length away from it. 

In the general context, it is not possible to determine the QNMs analytically, but in the regime $M\omega \ll 1$\footnote{For studying the gravitational wave echoes this is the most interesting regime since they are obtained from the ringdown signal after the merger, which initially is mostly dominated by the black hole-like QNMs and subsequently decays with time. The photon sphere, corresponding to the maxima of the potential, acts as a high-pass filter. This decreases the frequency content for each subsequent echo. Hence, at late times the low frequency approximation becomes more accurate\cite{Testa:2018bzd}.} one can determine the QNMs using analytical techniques as we will now demonstrate. We will follow the approach used in \cite{Starobinsky:1973aij,Maggio:2018ivz,Cardoso:2008kj} by matching the asymptotic expansion with the corresponding solution in the near horizon regime and hence solving for the discrete frequencies of the QNMs associated with the spacetime described in \ref{metric}. In brief, we solve the radial part of the scalar wave equation in the near horizon limit as well as in the asymptotic regime and then we match the solutions in an overlapping region defined as  $M\ll (r-r_{\rm h})\ll 1/\omega$, leading to the frequencies of the discrete tower of the QNMs.  

\subsection{Asymptotic solutions}

In this section, we will take the first step in obtaining analytical solutions for the QNMs, by determining the asymptotic behaviour of the solutions. For simplicity, we introduce the following definition, $\Delta\equiv r^2-2Mr-Q=(r-r_{\rm h})(r-r')$, where as mentioned earlier, $r'=M-\sqrt{M^2+Q}$ is used just as a notation, it does not represent any physical horizon. Thus plugging the spacetime metric presented in \ref{metric} in the wave equation, i.e., \ref{wave_eq_mod} and using the result $r^2\sqrt{fh}=\Delta$ we get,
\ba \la{wave_eq_hd}
\Delta\partial_r\big(\Delta \partial_r R_{\ell m}\big)+\big[r^4 \omega^2-\Delta \ell(\ell+1)\big]R_{\ell m}=0~.
\ea
Since we are interested in the asymptotic behaviour of this differential equation, we use the result that asymptotically $\Delta \sim r^{2}$ and thus \ref{wave_eq_hd} takes the following form,
\ba \la{wave_eq_infty}
\partial_r\big(r^2 \partial_r R_{\ell m}\big)+\big[r^2 \omega^2-\ell(\ell+1)\big]R_{\ell m}=0~.
\ea
This differential equation can be solved in terms of spherical Bessel functions as,
\ba \la{infty_soln}
R_{\ell m}=\frac{1}{\sqrt{r}}\left\{\alpha J_{\ell+1/2}(\omega r)+\beta J_{-\ell-1/2}(\omega r) \right\}~,
\ea
where $\alpha$ and $\beta$ are arbitrary constants of integration. This solution is in the asymptotic regime, while we want to match this solution with the near horizon solution in the intermediate regime. Thus we consider small $r$ limit of the above equation, yielding,
\ba \la{soln_infty_small}
R_{\ell m}\sim\alpha\frac{(\omega/2)^{\ell+1/2}}{\Gamma \big(\ell+3/2\big)}r^\ell+\beta \frac{(\omega/2)^{-\ell-1/2}}{\Gamma \big(-\ell+1/2\big)}r^{-\ell-1}~.
\ea
In order to arrive at the above expression we have used the following expansion of the Bessel function for small values of r, i.e., $J_{b}(r)=\{1/\Gamma(b+1)\}\left(r/2\right)^b$.

Along identical lines we can also determine the behaviour of the solution written down in \ref{infty_soln} for large values of the radial coordinate $r$, which yields,
\ba \la{soln_infty_large}
R_{\ell m}\sim\frac{1}{r}\sqrt{\frac{2}{\pi \omega}}\Big[\alpha \sin(\omega r-\ell \pi/2)+\beta \cos(\omega r+\ell \pi/2)\Big]~.
\nonumber\\
\ea
We will need both the small $r$ as well as large $r$ behaviour of the solution, as presented above, in the next section when we include appropriate boundary conditions and solve for the frequencies of the QNMs. 

\subsection{Near-horizon solutions}

As in the above section, we can perform a similar computation and write \ref{wave_eq_hd} in the near horizon limit, i.e., $r\sim r_{\rm h}$. In this context it is useful to define a new variable 
\ba
x=\frac{r-r_{\rm h}}{r_{\rm h}-r'};    
\qquad 
\partial_x=(r_{\rm h}-r')^{-1} \partial_r~.
\ea
Using which we can express the metric components, which effectively are determined by the quantity $\Delta$ alone, in terms of the newly defined variable $x$ as,
\ba
\Delta=(r-r_{\rm h})(r-r')= x(x+1)(r_{\rm h}-r')^2~.
\ea
Thus using this result as well as the near horizon approximation appropriately, from \ref{wave_eq_hd} we obtain the following form for the differential equation depicting radial evolution of the scalar field as,
\ba\la{near_horizon_eq}
\left(r_{\rm h}-r'\right)^{4}x(x+1)\partial_x\Big\{x(x+1)\partial_xR_{\ell m}\Big\}
+\Big\{\omega^2r_{\rm h}^4-(r_{\rm h}-r')^2x(x+1)\ell(\ell+1)\Big\}R_{\ell m}=0~.
\ea
Redefining the frequency $\omega$ appearing in the above expression as $\bar{\omega}\equiv \omega r_{\rm h}^2/(r_{\rm h}-r')$, the above differential equation, presented in \ref{near_horizon_eq}, can be re-expressed as,
\ba
 x^2(x+1)^2\partial_x^2R_{\ell m}+x(x+1)\Big(2x+1\Big)\partial_xR_{\ell m}
 +\Big\{\bar{\omega}^2-(x(x+1)\ell(\ell+1)\Big\}R_{\ell m}=0~.
\ea
At this stage it is instructive to define a new radial function $F_{\ell m}(r)$ such that, $R_{\ell m}(r)=(1+x)^{i\bar{\omega}}x^{-i\bar{\omega}}F_{\ell m}(r)$ and performing a change of variable $z\equiv 1-x$, \ref{near_horizon_eq} can be presented in the known hypergeometric form, which reads
\ba
z(1-z)F_{\ell m}''+\Big\{(1-2i\bar{\omega})-2z \Big\}F_{\ell m}'
 +\ell(\ell+1)F_{\ell m}=0~.
\ea
The solution of the above equation can be expressed in terms of hypergeometric functions and hence determine $F_{\ell m}(z)$ as a suitable linear combination of these hypergeometric functions. From which one can reverse the transformation and read of the scalar perturbation $R_{\ell m}$, which in terms of the variable $x$ takes the following form,
\ba \la{soln_near}
R_{\ell m}=(1+x)^{i\bar{\omega}}\Big[A x^{-i\bar{\omega}}{}_2F_{1}\left(-\ell,\ell+1,1-2i\bar{\omega},-x\right)
+B x^{i\bar{\omega}}{}_2F_1\left(-\ell+2i\bar{\omega},\ell+1+2i\bar{\omega},1+2i\bar{\omega},-x\right)\Big]~,
\ea
where $A$ and $B$ are arbitrary constants of integration. For small values of $x$, i.e., closer to the horizon we can expand the hypergeometric functions and obtain the following power law solution,
\ba \la{soln_near_near}
R_{\ell m}\sim Ax^{-i\bar{\omega}}+Bx^{i\bar{\omega}}
\ea
On the other hand, we need to obtain the behaviour of the near-horizon solution in the large $r$ regime, in order to match with the asymptotic solutions in the small radial distance limit, obtained in \ref{soln_infty_small}. For large values of $x$ (or, equivalently $r$) we can expand the hypergeometric functions as power series in the radial distance. Keeping the leading order terms in the expansion, and using suitable identities involving $\Gamma$ functions, we finally arrive at the following expression for the radial perturbation $R_{\ell m}$ associated with the scalar field,
\ba\la{soln_near_large}
R_{\ell m}\sim
\bigg(\frac{r}{r_{\rm h}-r'} \bigg)^\ell~\frac{\Gamma(2\ell+1)}{\Gamma(\ell+1)}\bigg[ A\frac{\Gamma(1-2i\bar{\omega})}{\Gamma(1+\ell-2i\bar{\omega})}+B\frac{\Gamma(1+2i\bar{\omega})}{\Gamma(\ell+1+2i\bar{\omega})}\bigg]
\nonumber\\
+
\bigg(\frac{r}{r_{\rm h}-r'} \bigg)^{-\ell-1}~\frac{\Gamma(-2\ell-1)}{\Gamma(-\ell)}\bigg[ A\frac{\Gamma(1-2i\bar{\omega})}{\Gamma(-\ell-2i\bar{\omega})}+B\frac{\Gamma(1+2i\bar{\omega})}{\Gamma(-\ell+2i\bar{\omega})}\bigg] \nonumber\\
\ea
Thus we have both the asymptotic, as well as near horizon solutions, along with their expansions in the intermediate overlapping region as well as in the asymptotic and near horizons regimes. We will now apply suitable boundary conditions on the asymptotic and near horizon form of the solutions to fix the arbitrary constants appearing in \ref{infty_soln} and \ref{soln_near} respectively, whose subsequent matching in the intermediate region will lead to an estimation of the frequencies of the QNMs.

\subsection{Boundary conditions and QNMs}

To obtain the QNMs we need to impose specific boundary conditions at the horizon and at asymptotic infinity. Since these boundary conditions are imposed on the in-going and out-going modes, it will be beneficial to pause for a while and understand the in-going and out-going modes. In this context, it would be helpful to introduce the tortoise coordinate, which for the static and spherically symmetric metric written down in \ref{eq1} can be calculated from the following differential relation $dr_{*}=\{1/\sqrt{f(r)h(r)}\}dr $. Performing the integration, the tortoise coordinate in this particular context of brane localized black hole with a negative tidal charge, yields,
\ba \la{tortoise}
r_{*}=\int\left( 1-\frac{2M}{r}-\frac{Q}{r^2}\right)^{-1}dr=\int \frac{r^2 dr}{(r-r_{\rm h})(r-r')}
\simeq\int\frac{r_{\rm h}^2~dx}{x(x+1)(r_h-r')}~.
\ea 
where in the last line we have used near horizon approximation. The near horizon approximation, is also manifested by the limit in which $x$ is small and hence the above integral yields
\ba \la{tor}
x\sim \exp\left\{r_{*}\left(\frac{r_h-r'}{r_h^2}\right)\right\}~.
\ea
Since at infinity, $r_{*}\sim r$, there is no point in doing this integral in that regime separately. As far as the boundary conditions are concerned, at infinity we impose the condition that only out-going waves are present. Hence we set the in-going part, i.e., coefficient of $\exp(-i\omega r)$ appearing in \ref{soln_infty_large} to zero and get the following condition
\ba
\beta=-i\alpha e^{i\pi \ell}~.
\ea
Similarly, near the apparent horizon, i.e., at $r=r_0\sim r_{\rm h}+\epsilon$, it is legitimate to introduce a reflective boundary condition with reflectivity $\mathcal{R}$\footnote{As black holes on the brane have a natural quantum interpretation \cite{Emparan:2002px} and hence following \cite{Oshita:2019sat} we should have used the Boltzmann reflectivity, but here for simplicity we are choosing a frequency independent reflectivity of the horizon to demonstrate the effect of the tidal charge on the ringdown spectrum.}. This yields, 
\be \la{near_boundary}
\frac{B}{A}x_0^{2i\bar{\omega}}=-\mathcal{R} 
\ee
where $x_{0}=x(r=r_{0})$, i.e., represents the value of the function $x(r)$ on the stretched horizon. Furthermore, note that for perfectly reflective boundary condition, we have $\mathcal{R}=1$ and it follows from \ref{near_boundary} that the field identically vanishes on the reflecting surface $r=r_{0}$. 

Thus using the boundary conditions at asymptotic infinity and near horizon regime, we have eliminated two unknown constants, appearing in the solution. However, there are still two more unknown constants present in the solutions. To fix these, we match the near horizon solution presented in \ref{soln_near_large} with the far region solution written down in \ref{soln_infty_small}, this yields,
\ba \la{matching}
\frac{B}{A}=-\prod_{n=1}^{\ell}\bigg[ \frac{n+2i\bar{\omega}}{n-2i\bar{\omega}}\bigg]
\bigg[\frac{1+2L(r_h-r')^{2\ell+1}\bar{\omega}\omega^{2\ell+1}\prod_{n=1}^{\ell}(n^2+4\bar{\omega}^2)}{1-2L(r_h-r')^{2\ell+1}\bar{\omega}\omega^{2\ell+1}\prod_{n=1}^{\ell}(n^2+4\bar{\omega}^2)}\bigg]~,
\ea
where the quantity $L$ appearing in the above equation is defined as
\be
L\equiv \frac{\pi \{\Gamma(\ell+1)\}^2}{2^{2\ell+2}\Gamma(\ell+3/2)\Gamma(2\ell+2)\Gamma(2\ell+1)\Gamma(\ell+1/2)}
\ee
As evident, from \ref{near_boundary} and \ref{matching} we can eliminate the ratio $(B/A)$ and obtain at $x_0$, i.e., on the stretched horizon, the following expression,
\ba \la{matching_r_0}
x_0^{2i\bar{\omega}}\prod_{n=1}^{\ell}\bigg[ \frac{n+2i\bar{\omega}}{n-2i\bar{\omega}}\bigg]
\bigg[\frac{1+2L(r_h-r')^{2\ell+1}\bar{\omega}\omega^{2\ell+1}\prod_{n=1}^{\ell}(n^2+4\bar{\omega}^2)}{1-2L(r_h-r')^{2\ell+1}\bar{\omega}\omega^{2\ell+1}\prod_{n=1}^{\ell}(n^2+4\bar{\omega}^2)}  \bigg]
=\mathcal{R}
\ea
Unfortunately, this equation cannot be solved in general using analytical methods in order to get the QNM frequency $\omega$, and we must resort to numerical techniques, which are discussed in the next section. However, under certain reasonable assumptions, this equation reduces to a much simpler form and it is possible to determine the frequencies of the QNMs analytically. First of all we have to work in the small frequency approximation, i.e., we will work in the limit of small $\omega$. If we further assume $M\omega\ll 1$ and the real part of $\omega$ is much larger than the imaginary part, then the above equation can be written down as, $\mathcal{R}=x_{0}^{2i\bar{\omega}}$. This can be re-expressed using \ref{tor} as
\ba
|\mathcal{R}|e^{i\left(\delta+2n\pi\right)}=\exp\left[2i\bar{\omega}r_{*}^{0}\left(r_{\rm h}-r'\right)/r^2_{\rm h}\right]~,
\ea
where $\delta$ is a phase factor depending on the details of the reflective membrane and $n$ is an integer. From this we can determine the real part of the quasi-normal mode frequency $\omega$ as 
\ba
\textrm{Re}(\omega)\sim \frac{\left(n\pi+\delta/2\right)}{r_{*}^{0}}
=\frac{2(n\pi+\delta/2)\sqrt{M^2+Q}}{r_{\rm h}^2\ln\left[\epsilon/\left(2\sqrt {M^2+Q}\right)\right]}~,
\ea
where we have expressed the term $r^0_*$ appearing in the above equation in terms of the black hole parameters and horizon radius as, 
\ba
r^{0}_{*}=\frac{r_{\rm h}^{2}}{2\sqrt{M^2+Q}}\ln \Big[\frac{\epsilon}{2\sqrt{M^2+Q}}\Big]~.
\ea
In order to determine the imaginary part of QNM frequency $\omega$ we can add a small imaginary part to the real part, i.e., $\omega\sim \textrm{Re}(\omega)+i\Omega$ and treat $\Omega$ in a perturbative manner. Thus the redefined frequency $\bar{\omega}$ becomes, $\bar{\omega}\sim \textrm{Re}(\bar{\omega})+i\sigma\Omega$, where $\sigma=r_{\rm h}^2/(r_{\rm h}-r')$. Performing the previous calculation with the small imaginary frequency thrown in we get from \ref{matching_r_0} the follwing relation,
\ba
x_0^{2i\textrm{Re}(\bar{\omega})}x_0^{-2\sigma\Omega}\prod_{n=1}^{\ell}
\bigg[\frac{n+2i\textrm{Re}(\bar{\omega})}{n-2i\textrm{Re}(\bar{\omega})}\bigg]
\bigg[\frac{1+2L(r_h-r')^{2\ell+1}\textrm{Re}(\bar{\omega})\omega^{2\ell+1}\prod_{n=1}^{\ell}\{n^2+4\textrm{Re}(\bar{\omega})^2\}}{1-2L(r_h-r')^{2\ell+1}\textrm{Re}(\bar{\omega})\omega^{2\ell+1}\prod_{n=1}^{\ell}\{n^2+4\textrm{Re}(\bar{\omega})^2\}}  \bigg]
=\mathcal{R}\nonumber
\\
\ea
This algebraic equation can also be solved analytically under the same set of assumptions as before and hence using the assumption for small $M\omega$ as well as $\bar{\omega}$, we get the following simplified form of the above equation,
\ba
x_{0}^{2\sigma\Omega}\sim
\frac{1+2L(r_h-r')^{2\ell+1}\textrm{Re}(\bar{\omega})\omega^{2\ell+1}\prod_{n=1}^{\ell}\{n^2+4\textrm{Re}(\bar{\omega})^2\}}{1-2L(r_h-r')^{2\ell+1}\textrm{Re}(\bar{\omega})\omega^{2\ell+1}\prod_{n=1}^{\ell}\{n^2+4\textrm{Re}(\bar{\omega})^2\}}
\frac{1}{\mathcal{R}}~. 
\ea
The imaginary part of $\omega$ can be immediately determined by solving the above equation, yielding,
\ba
&&\textrm{Im}(\omega)=\Omega\sim\frac{1}{2r^0_*}\ln \bigg[\frac{1+2L(r_h-r')^{2\ell+1}\textrm{Re}(\bar{\omega})\omega^{2\ell+1}\prod_{n=1}^{\ell}\{n^2+4\textrm{Re}(\bar{\omega}\}^2)}{1-2L(r_h-r')^{2\ell+1}\textrm{Re}(\bar{\omega})\omega^{2\ell+1}\prod_{n=1}^{\ell}\{n^2+4\textrm{Re}(\bar{\omega}\}^2)}\frac{1}{\mathcal{R}} \bigg]
\nonumber\\
&&=2L\textrm{Re}(\omega)~\left[\textrm{Re}(\omega)(r_h-r')\right]^{2\ell+1}\left(\frac{r_h^2}{\left(r_h-r'\right)r^0_*}\right)-\frac{\ln(\mathcal{R})}{2r_*^0}~.
\ea
where we have expanded the Logarithmic term in a power series and have kept the leading order term in the expansion.  We note that the real part of the QNM does not explicitly depend on the reflectivity of the membrane near the horizon but is sensitive to the position of the membrane. While the imaginary part depends both on the reflectivity, as well as on the location of the membrane, through the real part of the QNM frequency. As mentioned earlier, these are derived in the low frequency regime and hence is not applicable for any general frequency. To remedy this issue, we will determine the QNMs numerically in the next section.   
 
To summarize, we have solved the perturbation equation due to a scalar field in the spacetime described by \ref{metric} in two limits. First, in the asymptotic region, where the solutions are given by Bessel functions and in the near horizon regime, with hypergeometric functions. This brings in four unknown constants. Two of them got fixed by imposing the condition that at infinity the field modes are outgoing, while near the horizon there are ingoing and outgoing modes with some reflectivity. The other two constants are fixed by matching the large $r$ and near horizon solution in an intermediate region, which in turn yields an equation determining the quasi-normal modes. The most important and non-trivial step in the above analysis is the matching of the asymptotic solution with the near horizon solution, which becomes possible as the small $r$ behaviour of the asymptotic solution and the large $r$ behaviour of the near-horizon solution is identical. 
\section{Numerical analysis of the ringdown phase and echoes}\label{Sec_5}

\begin{figure}[h]
\centering
\includegraphics[width=12cm]{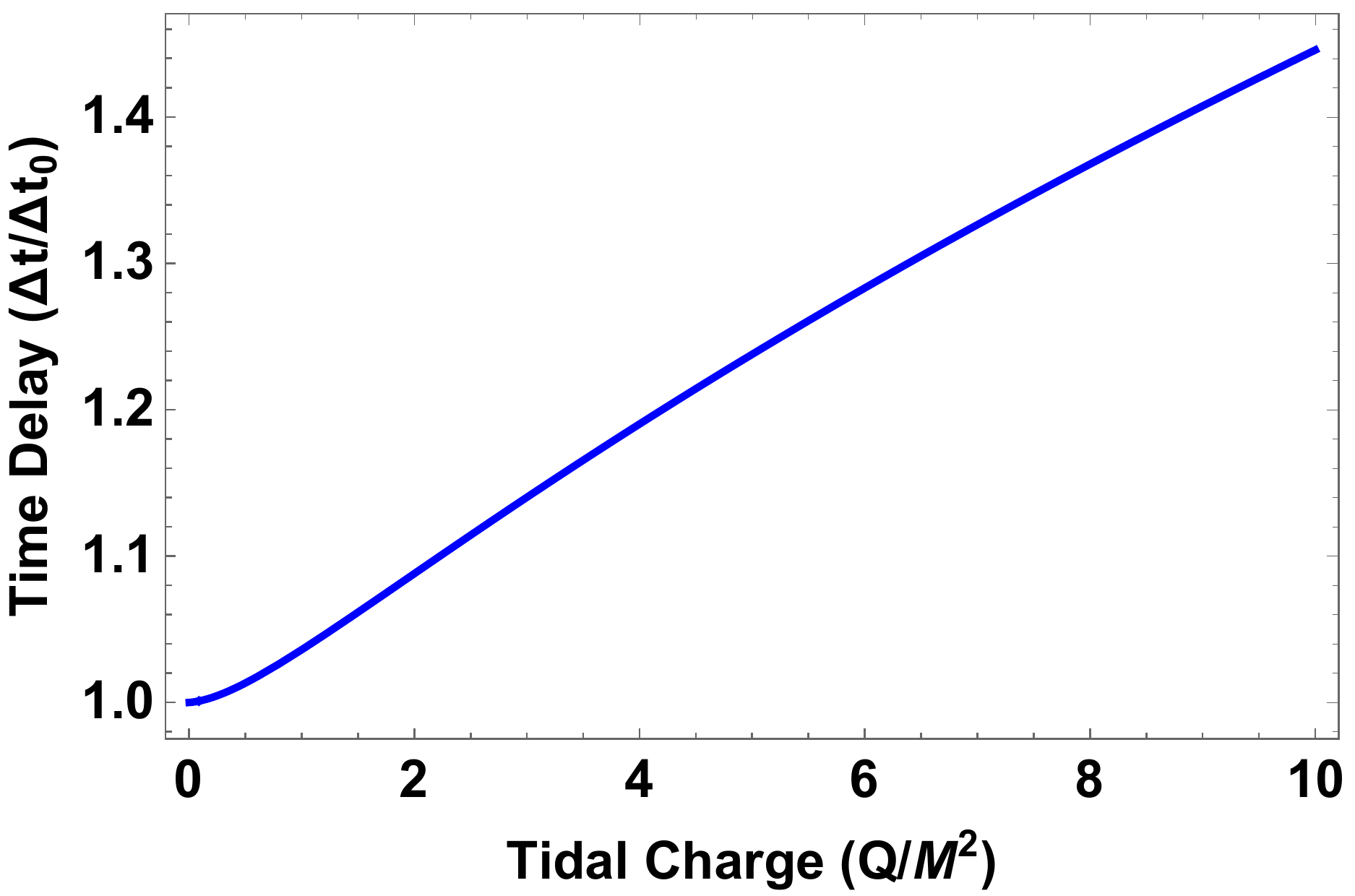}
\caption{From the analytic expression of the time delay $\Delta t$, leading to echo in the quasi-normal mode spectrum, as presented in \ref{echo_t}, we can see that $\Delta t$ depends on the tidal charge $Q$ as the position of the photon sphere as well as the horizon depends on it. Following which we have plotted the variation of the time delay $\Delta t$ with the tidal charge parameter $Q$, normalized to the time delay for $Q/M^2=0$. This explicitly demonstrates that the relative time delay increases with an increase in the tidal charge parameter $Q$.}\label{fig:time}
\end{figure}

As we assume a reflective boundary condition while computing the QNMs by solving \ref{near_boundary}, where the ingoing wave is (partially) reflected by some membrane close to the horizon, copies of the reflected wave reaches infinity after a time delay. This time delay can be calculated as the time it takes for the wave to travel in between the maxima of the potential (located at the photon circular orbit $r_{\rm ph}$) and the membrane close to the horizon (located at $r_0\sim r_{h}+\epsilon$). As mentioned earlier, such a reflective boundary condition originates from the quantum nature of brane localized black hole. Since the primary ringdown signal is coming through the potential and the secondary echoes comes after getting reflected from the potential and then from the stretched horizon at $r_{0}$. This induces a time delay $\Delta t$ in arrival of the secondary echoes, given as
\ba
\Delta t\simeq 2\int^{r_{\rm ph}}_{r_0}\frac{dr}{\sqrt{f(r)h(r)}}~.
\ea
For the metric given by \ref{metric}, the integrand is the same as the tortoise coordinate calculated in \ref{tortoise}, except for the specific limits used in this integral. Thus taking a cue from \ref{tortoise} we get,
\ba \label{echo_t}
&&\Delta t=2\int^{r_{\rm ph}}_{r_{\rm h}+\epsilon} \frac{r^2 dr}{(r-r_{\rm h})(r-r')}
\nonumber\\
&&=2r_{\rm ph}-2r_{\rm h}-2\epsilon+2\frac{r_{\rm h}^2}{r_{\rm h}-r'}\ln \bigg[ \frac{r_{\rm ph}-r_{\rm h}}{\epsilon}\bigg]
\nonumber\\
&&-2\frac{r'^2}{r_{\rm h}-r'}\ln \bigg[ \frac{r_{\rm ph}-r'}{r_{\rm h}-r'+\epsilon}\bigg]~.
\ea
It is interesting to note that the time delay is dependent on the value of the tidal charge $Q$ as $r_{\rm h}$, $r'$ and $r_{\rm ph}$ depends on $Q$. Hence the arrival of secondary echoes will depend on the value of $Q$. This is better illustrated in \ref{fig:time} where we have explicitly depicted how the relative echo time (time delay $\Delta t$, normalized by $\Delta t$ for vanishing tidal charge) changes with the change in the tidal charge $Q$. This will be verified later on through numerical analysis of the spectrum for quasi-normal modes.

Pushing forward, we solve for the scalar perturbation equation expressed in \ref{wave_eq_mod} numerically, assuming a reflective boundary condition at the horizon and obtain the QNMs. As the analytic approach is valid in the low frequency range only, we use numerical methods to obtain the echo spectrum for different values of $Q$ to demonstrate what kind of effect the tidal charge $Q$ has on the ringdown spectrum. If we assume that a primary signal is produced at the maxima of the potential, i.e., near the photon sphere, due to perturbation of the black hole, the echo waveform is given by the linear combination of the QNM's. This can be written as
\ba \la{waveform}
\Psi(t)=\sum^{\infty}_{n=-\infty}c_ne^{-i\omega_nt}
\ea
where the coefficients $c_{n}$ are the complex amplitude of the QNM's. For determining the amplitude, one can use various analytical templates   \cite{Mark:2017dnq,Testa:2018bzd} after taking the primary signal to be the same as the black holes fundamental QNM. Using the $\omega_n$ as we obtained by solving \ref{matching_r_0} numerically and choosing a suitable initial condition, one can obtain the waveform from \ref{waveform}. 
\begin{table}[h]
\begin{centering}
\begin{tabular}{|c|c|c|} 
\hline
\textbf{n} & \textbf{Numerical} & \textbf{Analytic}\\
\hline
\hline
0 & 0.023-$i~3.758\times 10^{-4}$ & 0.022-$i~3.663\times 10^{-4}$\\
1 & 0.069 -$i~3.755\times 10^{-4}$& 0.067-$i~3.66\times 10^{-4}$\\
2 & 0.115-$i~3.750\times 10^{-4}$& 0.119-$i~3.654\times 10^{-4}$\\
\hline
\end{tabular}
\par \end{centering}
\caption{Numerical and analytical estimations of the three lowest lying quasi-normal modes for the brane world black hole has been presented for $(Q/M^{2})=0.1$. As evident the numerical and analytical estimates match quite well. This shows the usefulness of the analytical approximations for computations of the quasi-normal modes.}
\label{tab:table1}
\end{table}

Having derived the waveform and hence the quasi-normal modes using numerical analysis, one can check the validity of the analytical approximations presented in the previous section,  by direct comparison. In particular, the validity of the matching of the asymptotic and near horizon solution, in the intermediate region, can also be tested by comparing with the numerical estimation for the quasi-normal modes. To see this explicitly, we have presented both the analytical and numerical estimations of the three lowest-lying quasi-normal modes in \ref{tab:table1}, which explicitly demonstrates the nice match between both the estimations. This suggests that the analytical approximations are good enough so that the analytical estimation of the quasi-normal modes matches with numerical results.

\begin{figure}
\centering
\includegraphics[height=14cm,width=8cm]{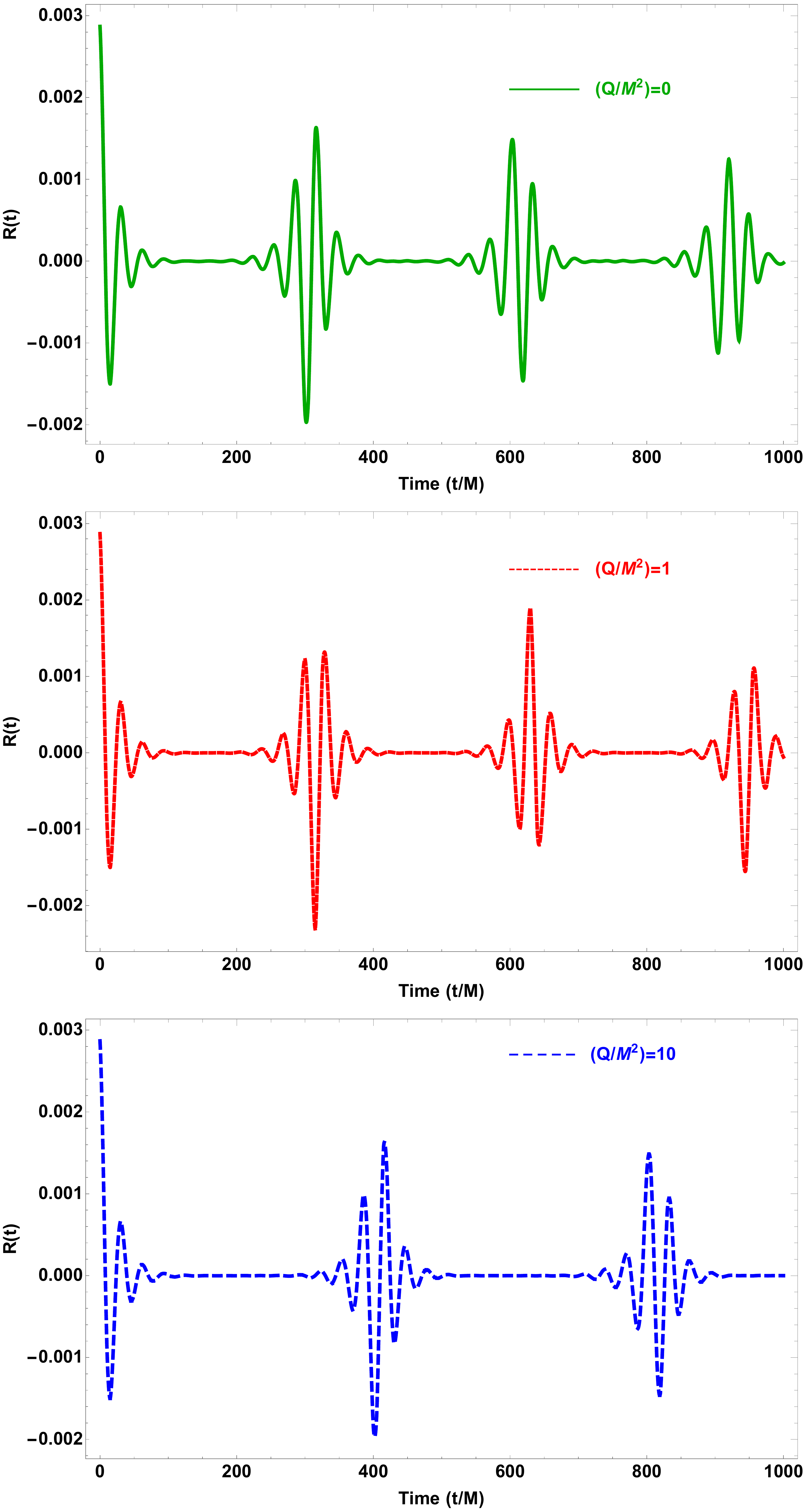}
\caption{The consecutive echoes along with the primary ringdown waveform is shown for three different values of the tidal charge, namely $(Q/M^{2})=0$, $(Q/M^{2})=1$ and $(Q/M^{2})=10$. For this particular numerical analysis we assumed $\epsilon \sim 10^{-18}\textrm{m}$, which follows from the bulk AdS curvature scale $\sim1\textrm{mm}$. As evident from the above figure, the echo time delay along with its amplitude is sensitive to the choice of the tidal charge parameter $Q$, in particular, the echo time delay increases with increase of the tidal charge parameter. Thus echo spectrum is another avenue to look for presence of extra dimensions and is intimately related to the size of the extra dimension. See text for more discussion.}\label{fig:Echo}
\end{figure}

The result of such an analysis and the resulting waveform has been presented in \ref{fig:Echo} for three different choices of the tidal charge parameter $Q$. We observe that the echo waveforms depict a significant departure from general relativistic prediction as the value of the tidal charge parameter is increased. In particular, with increase of the tidal charge $Q$, the echoes arrive at a later instant of time than their general relativistic counterparts. For example, if we consider the tidal charge parameter $Q/M^2\sim 1$, the echo arrival time would increase by $\sim 3\%$  (see \ref{fig:time}), which is roughly the level of precision for measurement of echo time at current instrumental sensitivities (e.g., Equation 6 in \cite{Abedi:2016hgu}).  

This provides a very distinct signature of the existence of extra dimensions in the context of black hole echoes, and could help to break the potential degeneracies between various parameters in the merger/ringdown signals. The physics behind such a difference can be understood along the following lines: As the negative tidal charge parameter $Q$ gets larger and larger, the effective potential experienced by the outgoing waves decreases in height, but is also further and further away from the stretched horizon. Thus the time delay increases with increasing $Q$. This results in the out of phase behaviour of the echoes. The difference in the time delay between two consecutive echo signals for different values of $Q$ can be also seen from the analytic expression \ref{echo_t} for the time delay which increases with the increase of the tidal charge $Q$ (also see \ref{fig:time}). Thus not only echoes are more natural in the braneworld scenario, but they also carry distinct signature of the existence of higher dimensions, which can be probed by considering the time delay of the echoes. 

Another point must be emphasized here, apparently, the theory presented above has two independent parameters $\epsilon$ and $Q$. This shows that there can be a potential degeneracy between these two parameters and hence it may not be possible to see the effect of the tidal charge unless some understanding of $\epsilon$ is present. This is not really the case for two separate reasons. First of all, as the tidal charge parameter gets modified, besides shifting of the maxima of the potential, the height of the potential also changes (see \ref{fig:potential}). Hence an increase in the tidal charge will not only change the echo time delay (which can also be changed by changing $\epsilon$) \emph{but} will also change the amplitude of the Echos. Such a change in amplitude can not brought in by $\epsilon$. Secondly, in the present context, $\epsilon$ crucially depends on the degrees of freedom of the boundary CFT, which in turn depends on the length scale of the AdS bulk. The AdS length scale is related to the horizon size into the bulk, which has non-trivial dependence on the tidal charge. Thus in this sense $\epsilon$ and $Q$ are really not independent, though a straightforward analytical relation does not exist.

Besides exploring the ringdown phase, a proper understanding of the inspiral phase is also important to provide a numerical estimate for the tidal charge parameter. In particular, it is important to develop the gravitational waveform, properly tuned with the effective gravitational field equations on the brane. However, calculating the gravitational waveforms of merging black holes in higher dimensions is currently beyond our scope. We do believe that it will be of utmost importance to address this issue, as future generations of gravitational wave detectors comes into the picture. In summary, our results provide the basic landscape on which we would like to improve upon in the future by further refining our analysis, e.g., including the inspiral part of black hole merger with appropriate wave forms.

\section{Discussion}

In the paradigm of AdS/CFT conjecture, the black holes localized on the brane are necessarily quantum corrected due to the coupling of the gravity and the CFT on the brane \cite{Emparan:2002px,Figueras_2011,Anderson_2005}. It is argued that there would be a difference between the event horizon and the apparent horizon of the black hole due to the backreaction of the CFT, which makes the black hole dynamical (e.g., the black hole will undergo Hawking radiation) \cite{Fabbri_2007}. Considering the dynamics of an evaporating black hole, one can show that the difference between these horizons would scale as $ l_{\rm p}^2$ and for evaporating black holes on the brane, this would be much larger and get multiplied by a factor of $N^2$, the number of degrees of freedom of the CFT on the brane \ref{difference}. The quantum effects on the horizon can have many non-trivial behaviours in contrast to those of a classical black hole, such as the presence of the {\it quantum ergosphere} and the back-reaction can make the horizon singular as well \cite{Anderson_2005,Gregory}. Without going into the model-dependent details of these speculations, it will be difficult to address the structure of the horizon in a four dimensional setting. However, it is clear that generically the boundary condition at the outer apparent horizon need not be purely ingoing. To quantify this, we assumed a reflective membrane, a Planck proper length away from the horizon. Thus, it naturally follows that the black holes localized on the brane must inherit such quantum corrections and as a consequence the natural boundary condition on the horizon may not be strictly ingoing. Thus, an open-minded reader may conclude that echoes may be natural bi-products of braneworld black holes.

Following this argument, we have analyzed the structure of the QNMs both analytically as well as numerically. In the analytical estimation, we have solved the scalar perturbation equation at the asymptotic region and the near horizon region. Subsequently, these solutions are matched at the intermediate region along with purely outgoing waves at infinity and partially reflecting wave at the stretched horizon. This provides an approximation for the real and imaginary parts of the QNM frequencies. It turns out that analytical estimates are possible only in the limit of small frequency. In a generic context, we have used appropriate numerical techniques to determine the QNM frequencies and have determined the waveform through numerical integration. This results in the appearance of echos in the ringdown signal.

We have also explicitly demonstrated the dependence of the echo spectrum on the tidal charge (see \ref{fig:Echo}). In particular, we have shown that the time scale of the echo, $\Delta t$, roughly given by
\begin{equation}
\Delta t \simeq \frac{2 \left(M+\sqrt{M^2+Q}\right)^2}{\sqrt{M^2+Q}} \ln\left(r_{\rm h} \over N \ell_{\rm p} \right),
\end{equation}
for non-spinning braneworld BH. 
$\Delta t$ increases as the tidal charge parameter, $Q$ is increased. This is easy to understand from \ref{fig:potential}, since with the increasing tidal charge, the maxima of potential goes to higher and higher radial distance compared to the $Q/M^2=0$ case. Therefore, the time scale for which the scalar modes are confined will increase, leading to increased time delay. Thus the presence of negative tidal charge will necessarily lead to an echo spectrum, which becomes out of phase as time progresses in comparison with the corresponding result from general relativity. In particular, for a tidal charge  $Q \sim M^2$, the arrival time of echoes can increase by $\sim 3\%$ in comparison to general relativity, which is at the current level of sensitivity of LIGO/Virgo measurements \cite{Abedi:2016hgu}. In combination with the merger/ringdown gravitational wave signal, this observation can provide smoking gun for the existence of negative tidal charge and hence of the extra spatial dimension, if the detection of echoes is confirmed in gravitational wave observations. Moreover, the location of the reflective membrane, which is also intimately related to the AdS radius, or equivalently, to the extent of the extra dimension, also has non-trivial implications for the echo time delay. Hence even the size of the extra dimension has non-trivial implications for the echo spectrum from braneworld black holes. Further, the case of braneworld black hole also gives a compelling answer to the question, why the horizon of a black hole should be reflective in nature.   

As future directions, it will be interesting to provide an observational constraint on the parameters for these higher dimensional black holes as that would require studying the full gravitational perturbation and knowing the extension of the brane localized black hole solution to the bulk, following \cite{Chakraborty:2017qve,Maartens:2010ar,Seahra:2004fg}. Extension for rotating brane localized black holes is also an interesting extension of this work and falls within the scope of a future application. Understanding the case of rotating black holes is necessary in order to make predictions about astrophysical black holes, which will be important to connect with the present day observations. A recent work has shown the presence of echoes for wormholes in the braneworld scenario  \cite{Bronnikov:2019sbx} which we wanted to explore as well. In addition, how the quantum corrections due to presence of CFT on the brane will affect the black hole solution considered here will be another interesting direction of exploration and will be reported elsewhere.  

\section*{Acknowledgments}

We would like to thank Naritaka Oshita for helping us with the numerical section and numerous stimulating discussions. Research of SC is funded by the INSPIRE Faculty fellowship Sir James Lougheed Award of Distinction Department of Science and Technology, Government of India (Reg. No. DST/INSPIRE/04/2018/000893). This work was partially supported by the University of Waterloo, Natural Sciences and Engineering Research Council of Canada (NSERC), the Perimeter Institute for Theoretical Physics. Research at the Perimeter Institute is supported by the Government of Canada through In- dustry Canada, and by the Province of Ontario through the Ministry of Research and Innovation.
\bibliography{QNM}

\providecommand{\href}[2]{#2}\begingroup\raggedright\begin{thebibliography}{10}

\bibitem{Abbott:2016blz}
{\bfseries LIGO Scientific, Virgo} Collaboration, B.~P. Abbott {\em et~al.},
  ``{Observation of Gravitational Waves from a Binary Black Hole Merger},''
  \href{http://dx.doi.org/10.1103/PhysRevLett.116.061102}{{\em Phys. Rev.
  Lett.} {\bfseries 116} no.~6, (2016) 061102},
\href{http://arxiv.org/abs/1602.03837}{{\ttfamily arXiv:1602.03837 [gr-qc]}}.

\bibitem{TheLIGOScientific:2016pea}
{\bfseries Virgo, LIGO Scientific} Collaboration, B.~P. Abbott {\em et~al.},
  ``{Binary Black Hole Mergers in the first Advanced LIGO Observing Run},''
  \href{http://dx.doi.org/10.1103/PhysRevX.6.041015}{{\em Phys. Rev.}
  {\bfseries X6} no.~4, (2016) 041015},
\href{http://arxiv.org/abs/1606.04856}{{\ttfamily arXiv:1606.04856 [gr-qc]}}.

\bibitem{Abbott:2016nmj}
{\bfseries Virgo, LIGO Scientific} Collaboration, B.~P. Abbott {\em et~al.},
  ``{GW151226: Observation of Gravitational Waves from a 22-Solar-Mass Binary
  Black Hole Coalescence},''
  \href{http://dx.doi.org/10.1103/PhysRevLett.116.241103}{{\em Phys. Rev.
  Lett.} {\bfseries 116} no.~24, (2016) 241103},
\href{http://arxiv.org/abs/1606.04855}{{\ttfamily arXiv:1606.04855 [gr-qc]}}.

\bibitem{TheLIGOScientific:2016src}
{\bfseries Virgo, LIGO Scientific} Collaboration, B.~P. Abbott {\em et~al.},
  ``{Tests of general relativity with GW150914},''
  \href{http://dx.doi.org/10.1103/PhysRevLett.116.221101}{{\em Phys. Rev.
  Lett.} {\bfseries 116} no.~22, (2016) 221101},
\href{http://arxiv.org/abs/1602.03841}{{\ttfamily arXiv:1602.03841 [gr-qc]}}.

\bibitem{Abbott:2017vtc}
{\bfseries VIRGO, LIGO Scientific} Collaboration, B.~P. Abbott {\em et~al.},
  ``{GW170104: Observation of a 50-Solar-Mass Binary Black Hole Coalescence at
  Redshift 0.2},'' \href{http://dx.doi.org/10.1103/PhysRevLett.118.221101}{{\em
  Phys. Rev. Lett.} {\bfseries 118} no.~22, (2017) 221101},
\href{http://arxiv.org/abs/1706.01812}{{\ttfamily arXiv:1706.01812 [gr-qc]}}.

\bibitem{PhysRevLett.116.171101}
V.~Cardoso, E.~Franzin, and P.~Pani, ``Is the gravitational-wave ringdown a
  probe of the event horizon?,''
  \href{http://dx.doi.org/10.1103/PhysRevLett.116.171101}{{\em Phys. Rev.
  Lett.} {\bfseries 116} (Apr, 2016) 171101}.
  \url{https://link.aps.org/doi/10.1103/PhysRevLett.116.171101}.

\bibitem{Kokkotas:1999bd}
K.~D. Kokkotas and B.~G. Schmidt, ``{Quasinormal modes of stars and black
  holes},'' \href{http://dx.doi.org/10.12942/lrr-1999-2}{{\em Living Rev. Rel.}
  {\bfseries 2} (1999) 2},
\href{http://arxiv.org/abs/gr-qc/9909058}{{\ttfamily arXiv:gr-qc/9909058
  [gr-qc]}}.

\bibitem{Mathur:2005zp}
S.~D. Mathur, ``{The Fuzzball proposal for black holes: An Elementary
  review},'' \href{http://dx.doi.org/10.1002/prop.200410203}{{\em Fortsch.
  Phys.} {\bfseries 53} (2005) 793--827},
\href{http://arxiv.org/abs/hep-th/0502050}{{\ttfamily arXiv:hep-th/0502050
  [hep-th]}}.

\bibitem{Almheiri:2012rt}
A.~Almheiri, D.~Marolf, J.~Polchinski, and J.~Sully, ``{Black Holes:
  Complementarity or Firewalls?},''
  \href{http://dx.doi.org/10.1007/JHEP02(2013)062}{{\em JHEP} {\bfseries 02}
  (2013) 062},
\href{http://arxiv.org/abs/1207.3123}{{\ttfamily arXiv:1207.3123 [hep-th]}}.

\bibitem{Bueno:2017hyj}
P.~Bueno, P.~A. Cano, F.~Goelen, T.~Hertog, and B.~Vercnocke, ``{Echoes of
  Kerr-like wormholes},''
  \href{http://dx.doi.org/10.1103/PhysRevD.97.024040}{{\em Phys. Rev.}
  {\bfseries D97} no.~2, (2018) 024040},
\href{http://arxiv.org/abs/1711.00391}{{\ttfamily arXiv:1711.00391 [gr-qc]}}.

\bibitem{Oshita:2019sat}
N.~Oshita, Q.~Wang, and N.~Afshordi, ``{On Reflectivity of Quantum Black Hole
  Horizons},''
\href{http://arxiv.org/abs/1905.00464}{{\ttfamily arXiv:1905.00464 [hep-th]}}.

\bibitem{PhysRevD.94.084031}
V.~Cardoso, S.~Hopper, C.~F.~B. Macedo, C.~Palenzuela, and P.~Pani,
  ``Gravitational-wave signatures of exotic compact objects and of quantum
  corrections at the horizon scale,''
  \href{http://dx.doi.org/10.1103/PhysRevD.94.084031}{{\em Phys. Rev. D}
  {\bfseries 94} (Oct, 2016) 084031}.
  \url{https://link.aps.org/doi/10.1103/PhysRevD.94.084031}.

\bibitem{Abedi:2016hgu}
J.~Abedi, H.~Dykaar, and N.~Afshordi, ``{Echoes from the Abyss: Tentative
  evidence for Planck-scale structure at black hole horizons},''
  \href{http://dx.doi.org/10.1103/PhysRevD.96.082004}{{\em Phys. Rev.}
  {\bfseries D96} no.~8, (2017) 082004},
\href{http://arxiv.org/abs/1612.00266}{{\ttfamily arXiv:1612.00266 [gr-qc]}}.

\bibitem{Oshita:2018fqu}
N.~Oshita and N.~Afshordi, ``{Probing microstructure of black hole spacetimes
  with gravitational wave echoes},''
  \href{http://dx.doi.org/10.1103/PhysRevD.99.044002}{{\em Phys. Rev.}
  {\bfseries D99} no.~4, (2019) 044002},
\href{http://arxiv.org/abs/1807.10287}{{\ttfamily arXiv:1807.10287 [gr-qc]}}.

\bibitem{Wang:2019rcf}
Q.~Wang, N.~Oshita, and N.~Afshordi, ``{Echoes from Quantum Black Holes},''
\href{http://arxiv.org/abs/1905.00446}{{\ttfamily arXiv:1905.00446 [gr-qc]}}.

\bibitem{Wang_2018}
Q.~Wang and N.~Afshordi, ``Black hole echology: The observer's manual,''
  \href{http://dx.doi.org/10.1103/physrevd.97.124044}{{\em Physical Review D}
  {\bfseries 97} no.~12, (Jun, 2018) }.
  \url{http://dx.doi.org/10.1103/PhysRevD.97.124044}.

\bibitem{Conklin:2017lwb}
R.~S. Conklin, B.~Holdom, and J.~Ren, ``{Gravitational wave echoes through new
  windows},'' \href{http://dx.doi.org/10.1103/PhysRevD.98.044021}{{\em Phys.
  Rev.} {\bfseries D98} no.~4, (2018) 044021},
\href{http://arxiv.org/abs/1712.06517}{{\ttfamily arXiv:1712.06517 [gr-qc]}}.

\bibitem{abedi2018echoes}
J.~Abedi and N.~Afshordi, ``Echoes from the abyss: A highly spinning black hole
  remnant for the binary neutron star merger gw170817,'' 2018.

\bibitem{Conklin:2019fcs}
R.~S. Conklin and B.~Holdom, ``{Gravitational Wave "Echo" Spectra},''
\href{http://arxiv.org/abs/1905.09370}{{\ttfamily arXiv:1905.09370 [gr-qc]}}.

\bibitem{Holdom:2019bdv}
B.~Holdom, ``{Not quite black holes at LIGO},''
\href{http://arxiv.org/abs/1909.11801}{{\ttfamily arXiv:1909.11801 [gr-qc]}}.

\bibitem{Westerweck:2017hus}
J.~Westerweck, A.~Nielsen, O.~Fischer-Birnholtz, M.~Cabero, C.~Capano, T.~Dent,
  B.~Krishnan, G.~Meadors, and A.~H. Nitz, ``{Low significance of evidence for
  black hole echoes in gravitational wave data},''
  \href{http://dx.doi.org/10.1103/PhysRevD.97.124037}{{\em Phys. Rev.}
  {\bfseries D97} no.~12, (2018) 124037},
\href{http://arxiv.org/abs/1712.09966}{{\ttfamily arXiv:1712.09966 [gr-qc]}}.

\bibitem{Tsang:2019zra}
K.~W. Tsang, A.~Ghosh, A.~Samajdar, K.~Chatziioannou, S.~Mastrogiovanni,
  M.~Agathos, and C.~Van Den~Broeck, ``{A morphology-independent search for
  gravitational wave echoes in data from the first and second observing runs of
  Advanced LIGO and Advanced Virgo},''
\href{http://arxiv.org/abs/1906.11168}{{\ttfamily arXiv:1906.11168 [gr-qc]}}.

\bibitem{Salemi:2019uea}
F.~Salemi, E.~Milotti, G.~A. Prodi, G.~Vedovato, C.~Lazzaro, S.~Tiwari,
  S.~Vinciguerra, M.~Drago, and S.~Klimenko, ``{Wider look at the
  gravitational-wave transients from GWTC-1 using an unmodeled reconstruction
  method},'' \href{http://dx.doi.org/10.1103/PhysRevD.100.042003}{{\em Phys.
  Rev.} {\bfseries D100} no.~4, (2019) 042003},
\href{http://arxiv.org/abs/1905.09260}{{\ttfamily arXiv:1905.09260 [gr-qc]}}.

\bibitem{Uchikata:2019frs}
N.~Uchikata, H.~Nakano, T.~Narikawa, N.~Sago, H.~Tagoshi, and T.~Tanaka,
  ``{Searching for black hole echoes from the LIGO-Virgo Catalog GWTC-1},''
  \href{http://dx.doi.org/10.1103/PhysRevD.100.062006}{{\em Phys. Rev.}
  {\bfseries D100} no.~6, (2019) 062006},
\href{http://arxiv.org/abs/1906.00838}{{\ttfamily arXiv:1906.00838 [gr-qc]}}.

\bibitem{Randall:1999ee}
L.~Randall and R.~Sundrum, ``{A Large mass hierarchy from a small extra
  dimension},'' \href{http://dx.doi.org/10.1103/PhysRevLett.83.3370}{{\em Phys.
  Rev. Lett.} {\bfseries 83} (1999) 3370--3373},
\href{http://arxiv.org/abs/hep-ph/9905221}{{\ttfamily arXiv:hep-ph/9905221
  [hep-ph]}}.

\bibitem{Randall:1999vf}
L.~Randall and R.~Sundrum, ``{An Alternative to compactification},''
  \href{http://dx.doi.org/10.1103/PhysRevLett.83.4690}{{\em Phys. Rev. Lett.}
  {\bfseries 83} (1999) 4690--4693},
\href{http://arxiv.org/abs/hep-th/9906064}{{\ttfamily arXiv:hep-th/9906064
  [hep-th]}}.

\bibitem{Maartens:2010ar}
R.~Maartens and K.~Koyama, ``{Brane-World Gravity},''
  \href{http://dx.doi.org/10.12942/lrr-2010-5}{{\em Living Rev. Rel.}
  {\bfseries 13} (2010) 5},
\href{http://arxiv.org/abs/1004.3962}{{\ttfamily arXiv:1004.3962 [hep-th]}}.

\bibitem{Shiromizu:1999wj}
T.~Shiromizu, K.-i. Maeda, and M.~Sasaki, ``{The Einstein equation on the
  3-brane world},'' \href{http://dx.doi.org/10.1103/PhysRevD.62.024012}{{\em
  Phys. Rev.} {\bfseries D62} (2000) 024012},
\href{http://arxiv.org/abs/gr-qc/9910076}{{\ttfamily arXiv:gr-qc/9910076
  [gr-qc]}}.

\bibitem{Harko:2004ui}
T.~Harko and M.~K. Mak, ``{Vacuum solutions of the gravitational field
  equations in the brane world model},''
  \href{http://dx.doi.org/10.1103/PhysRevD.69.064020}{{\em Phys. Rev.}
  {\bfseries D69} (2004) 064020},
\href{http://arxiv.org/abs/gr-qc/0401049}{{\ttfamily arXiv:gr-qc/0401049
  [gr-qc]}}.

\bibitem{Aliev:2005bi}
A.~N. Aliev and A.~E. Gumrukcuoglu, ``{Charged rotating black holes on a
  3-brane},'' \href{http://dx.doi.org/10.1103/PhysRevD.71.104027}{{\em Phys.
  Rev.} {\bfseries D71} (2005) 104027},
\href{http://arxiv.org/abs/hep-th/0502223}{{\ttfamily arXiv:hep-th/0502223
  [hep-th]}}.

\bibitem{Chakraborty:2015taq}
S.~Chakraborty and S.~SenGupta, ``{Spherically symmetric brane in a bulk of
  $f(R)$ and Gauss--Bonnet gravity},''
  \href{http://dx.doi.org/10.1088/0264-9381/33/22/225001}{{\em Class. Quant.
  Grav.} {\bfseries 33} no.~22, (2016) 225001},
\href{http://arxiv.org/abs/1510.01953}{{\ttfamily arXiv:1510.01953 [gr-qc]}}.

\bibitem{Chakraborty:2014xla}
S.~Chakraborty and S.~SenGupta, ``{Spherically symmetric brane spacetime with
  bulk $f(\mathcal {R})$ gravity},''
  \href{http://dx.doi.org/10.1140/epjc/s10052-014-3234-3}{{\em Eur. Phys. J.}
  {\bfseries C75} no.~1, (2015) 11},
\href{http://arxiv.org/abs/1409.4115}{{\ttfamily arXiv:1409.4115 [gr-qc]}}.

\bibitem{Dadhich:2000am}
N.~Dadhich, R.~Maartens, P.~Papadopoulos, and V.~Rezania, ``{Black holes on the
  brane},'' \href{http://dx.doi.org/10.1016/S0370-2693(00)00798-X}{{\em Phys.
  Lett.} {\bfseries B487} (2000) 1--6},
\href{http://arxiv.org/abs/hep-th/0003061}{{\ttfamily arXiv:hep-th/0003061
  [hep-th]}}.

\bibitem{Banerjee:2019nnj}
I.~Banerjee, S.~Chakraborty, and S.~SenGupta, ``{Silhouette of M87*: A New
  Window to Peek into the World of Hidden Dimensions},''
\href{http://arxiv.org/abs/1909.09385}{{\ttfamily arXiv:1909.09385 [gr-qc]}}.

\bibitem{Banerjee:2019sae}
I.~Banerjee, S.~Chakraborty, and S.~SenGupta, ``{Decoding signatures of extra
  dimensions and estimating spin of quasars from the continuum spectrum},''
  \href{http://dx.doi.org/10.1103/PhysRevD.100.044045}{{\em Phys. Rev.}
  {\bfseries D100} no.~4, (2019) 044045},
\href{http://arxiv.org/abs/1905.08043}{{\ttfamily arXiv:1905.08043 [gr-qc]}}.

\bibitem{Banerjee:2017hzw}
I.~Banerjee, S.~Chakraborty, and S.~SenGupta, ``{Excavating black hole
  continuum spectrum: Possible signatures of scalar hairs and of higher
  dimensions},'' \href{http://dx.doi.org/10.1103/PhysRevD.96.084035}{{\em Phys.
  Rev.} {\bfseries D96} no.~8, (2017) 084035},
\href{http://arxiv.org/abs/1707.04494}{{\ttfamily arXiv:1707.04494 [gr-qc]}}.

\bibitem{Vagnozzi:2019apd}
S.~Vagnozzi and L.~Visinelli, ``{Hunting for extra dimensions in the shadow of
  M87*},'' \href{http://dx.doi.org/10.1103/PhysRevD.100.024020}{{\em Phys.
  Rev.} {\bfseries D100} no.~2, (2019) 024020},
\href{http://arxiv.org/abs/1905.12421}{{\ttfamily arXiv:1905.12421 [gr-qc]}}.

\bibitem{Chakravarti:2019aup}
K.~Chakravarti, S.~Chakraborty, K.~S. Phukon, S.~Bose, and S.~SenGupta,
  ``{Constraining extra-spatial dimensions with observations of GW170817},''
\href{http://arxiv.org/abs/1903.10159}{{\ttfamily arXiv:1903.10159 [gr-qc]}}.

\bibitem{Chakravarti:2018vlt}
K.~Chakravarti, S.~Chakraborty, S.~Bose, and S.~SenGupta, ``{Tidal Love numbers
  of black holes and neutron stars in the presence of higher dimensions:
  Implications of GW170817},''
  \href{http://dx.doi.org/10.1103/PhysRevD.99.024036}{{\em Phys. Rev.}
  {\bfseries D99} no.~2, (2019) 024036},
\href{http://arxiv.org/abs/1811.11364}{{\ttfamily arXiv:1811.11364 [gr-qc]}}.

\bibitem{Chakraborty:2017qve}
S.~Chakraborty, K.~Chakravarti, S.~Bose, and S.~SenGupta, ``{Signatures of
  extra dimensions in gravitational waves from black hole quasinormal modes},''
  \href{http://dx.doi.org/10.1103/PhysRevD.97.104053}{{\em Phys. Rev.}
  {\bfseries D97} no.~10, (2018) 104053},
\href{http://arxiv.org/abs/1710.05188}{{\ttfamily arXiv:1710.05188 [gr-qc]}}.

\bibitem{Visinelli:2017bny}
L.~Visinelli, N.~Bolis, and S.~Vagnozzi, ``{Brane-world extra dimensions in
  light of GW170817},''
  \href{http://dx.doi.org/10.1103/PhysRevD.97.064039}{{\em Phys. Rev.}
  {\bfseries D97} no.~6, (2018) 064039},
\href{http://arxiv.org/abs/1711.06628}{{\ttfamily arXiv:1711.06628 [gr-qc]}}.

\bibitem{Rahman:2018oso}
M.~Rahman, S.~Chakraborty, S.~SenGupta, and A.~A. Sen, ``{Fate of Strong Cosmic
  Censorship Conjecture in Presence of Higher Spacetime Dimensions},''
  \href{http://dx.doi.org/10.1007/JHEP03(2019)178}{{\em JHEP} {\bfseries 03}
  (2019) 178},
\href{http://arxiv.org/abs/1811.08538}{{\ttfamily arXiv:1811.08538 [gr-qc]}}.

\bibitem{Mishra:2019ged}
A.~K. Mishra and S.~Chakraborty, ``{Strong Cosmic Censorship in higher
  curvature gravity},''
\href{http://arxiv.org/abs/1911.09855}{{\ttfamily arXiv:1911.09855 [gr-qc]}}.

\bibitem{Chamblin_2000}
A.~Chamblin, S.~W. Hawking, and H.~S. Reall, ``Brane-world black holes,''
  \href{http://dx.doi.org/10.1103/physrevd.61.065007}{{\em Physical Review D}
  {\bfseries 61} no.~6, (Feb, 2000) }.
  \url{http://dx.doi.org/10.1103/PhysRevD.61.065007}.

\bibitem{Bruni_2001}
M.~Bruni, C.~Germani, and R.~Maartens, ``Gravitational collapse on the brane: A
  no-go theorem,'' \href{http://dx.doi.org/10.1103/physrevlett.87.231302}{{\em
  Physical Review Letters} {\bfseries 87} no.~23, (Nov, 2001) }.
  \url{http://dx.doi.org/10.1103/PhysRevLett.87.231302}.

\bibitem{Chamblin_2001}
A.~Chamblin, H.~S. Reall, H.-a. Shinkai, and T.~Shiromizu, ``Charged
  brane-world black holes,''
  \href{http://dx.doi.org/10.1103/physrevd.63.064015}{{\em Physical Review D}
  {\bfseries 63} no.~6, (Feb, 2001) }.
  \url{http://dx.doi.org/10.1103/PhysRevD.63.064015}.

\bibitem{Kanti_2002}
P.~Kanti and K.~Tamvakis, ``Quest for localized 4d black holes in brane
  worlds,'' \href{http://dx.doi.org/10.1103/physrevd.65.084010}{{\em Physical
  Review D} {\bfseries 65} no.~8, (Mar, 2002) }.
  \url{http://dx.doi.org/10.1103/PhysRevD.65.084010}.

\bibitem{Emparan_2000}
R.~Emparan, G.~T. Horowitz, and R.~C. Myers, ``Exact description of black holes
  on branes,'' \href{http://dx.doi.org/10.1088/1126-6708/2000/01/007}{{\em
  Journal of High Energy Physics} {\bfseries 2000} no.~01, (Jan, 2000)
  007--007}. \url{http://dx.doi.org/10.1088/1126-6708/2000/01/007}.

\bibitem{Gregory}
R.~Gregory, ``Braneworld black holes,''
  \href{http://dx.doi.org/10.1007/978-3-540-88460-6_7}{{\em Lecture Notes in
  Physics} 259--298}. \url{http://dx.doi.org/10.1007/978-3-540-88460-6_7}.

\bibitem{Emparan:2002px}
R.~Emparan, A.~Fabbri, and N.~Kaloper, ``{Quantum black holes as holograms in
  AdS brane worlds},''
  \href{http://dx.doi.org/10.1088/1126-6708/2002/08/043}{{\em JHEP} {\bfseries
  08} (2002) 043},
\href{http://arxiv.org/abs/hep-th/0206155}{{\ttfamily arXiv:hep-th/0206155
  [hep-th]}}.

\bibitem{Anderson_2005}
P.~R. Anderson, R.~Balbinot, and A.~Fabbri, ``Cutoff anti--de sitter
  space/conformal-field-theory duality and the quest for braneworld black
  holes,'' \href{http://dx.doi.org/10.1103/physrevlett.94.061301}{{\em Physical
  Review Letters} {\bfseries 94} no.~6, (Feb, 2005) }.
  \url{http://dx.doi.org/10.1103/PhysRevLett.94.061301}.

\bibitem{Fabbri_2007}
A.~Fabbri and G.~P. Procopio, ``Quantum effects in black holes from the
  schwarzschild black string?,''
  \href{http://dx.doi.org/10.1088/0264-9381/24/22/003}{{\em Classical and
  Quantum Gravity} {\bfseries 24} no.~22, (Oct, 2007) 5371--5382}.
  \url{http://dx.doi.org/10.1088/0264-9381/24/22/003}.

\bibitem{Gregory_1993}
R.~Gregory and R.~Laflamme, ``Black strings andp-branes are unstable,''
  \href{http://dx.doi.org/10.1103/physrevlett.70.2837}{{\em Physical Review
  Letters} {\bfseries 70} no.~19, (May, 1993) 2837--2840}.
  \url{http://dx.doi.org/10.1103/PhysRevLett.70.2837}.

\bibitem{Gregory_2000}
R.~Gregory, ``Black string instabilities in anti-de sitter space,''
  \href{http://dx.doi.org/10.1088/0264-9381/17/18/103}{{\em Classical and
  Quantum Gravity} {\bfseries 17} no.~18, (Aug, 2000) L125--L131}.
  \url{http://dx.doi.org/10.1088/0264-9381/17/18/103}.

\bibitem{Bhattacharya:2016naa}
S.~Bhattacharya and S.~Chakraborty, ``{Constraining some Horndeski gravity
  theories},'' \href{http://dx.doi.org/10.1103/PhysRevD.95.044037}{{\em Phys.
  Rev.} {\bfseries D95} no.~4, (2017) 044037},
\href{http://arxiv.org/abs/1607.03693}{{\ttfamily arXiv:1607.03693 [gr-qc]}}.

\bibitem{Garriga_2000}
J.~Garriga and T.~Tanaka, ``Gravity in the randall-sundrum brane world,''
  \href{http://dx.doi.org/10.1103/physrevlett.84.2778}{{\em Physical Review
  Letters} {\bfseries 84} no.~13, (Mar, 2000) 2778--2781}.
  \url{http://dx.doi.org/10.1103/PhysRevLett.84.2778}.

\bibitem{Duff:2000mt}
M.~J. Duff and J.~T. Liu, ``{Complementarity of the Maldacena and
  Randall-Sundrum pictures},''
  \href{http://dx.doi.org/10.1088/0264-9381/18/16/310,
  10.1103/PhysRevLett.85.2052}{{\em Class. Quant. Grav.} {\bfseries 18} (2001)
  3207--3214}, \href{http://arxiv.org/abs/hep-th/0003237}{{\ttfamily
  arXiv:hep-th/0003237 [hep-th]}}.
[Phys. Rev. Lett.85,2052(2000)].

\bibitem{Emparan_2003}
R.~Emparan, J.~Garc{\'\i}a-Bellido, and N.~Kaloper, ``Black hole astrophysics
  in ads braneworlds,''
  \href{http://dx.doi.org/10.1088/1126-6708/2003/01/079}{{\em Journal of High
  Energy Physics} {\bfseries 2003} no.~01, (Jan, 2003) 079--079}.
  \url{http://dx.doi.org/10.1088/1126-6708/2003/01/079}.

\bibitem{Toshmatov:2016bsb}
B.~Toshmatov, Z.~Stuchl{\'\i}k, J.~Schee, and B.~Ahmedov, ``{Quasinormal
  frequencies of black hole in the braneworld},''
  \href{http://dx.doi.org/10.1103/PhysRevD.93.124017}{{\em Phys. Rev.}
  {\bfseries D93} no.~12, (2016) 124017},
\href{http://arxiv.org/abs/1605.02058}{{\ttfamily arXiv:1605.02058 [gr-qc]}}.

\bibitem{Testa:2018bzd}
A.~Testa and P.~Pani, ``{Analytical template for gravitational-wave echoes:
  signal characterization and prospects of detection with current and future
  interferometers},'' \href{http://dx.doi.org/10.1103/PhysRevD.98.044018}{{\em
  Phys. Rev.} {\bfseries D98} no.~4, (2018) 044018},
\href{http://arxiv.org/abs/1806.04253}{{\ttfamily arXiv:1806.04253 [gr-qc]}}.

\bibitem{Starobinsky:1973aij}
A.~A. Starobinsky, ``{Amplification of waves reflected from a rotating "black
  hole".},'' {\em Sov. Phys. JETP} {\bfseries 37} no.~1, (1973) 28--32.
[Zh. Eksp. Teor. Fiz.64,48(1973)].

\bibitem{Maggio:2018ivz}
E.~Maggio, V.~Cardoso, S.~R. Dolan, and P.~Pani, ``{Ergoregion instability of
  exotic compact objects: electromagnetic and gravitational perturbations and
  the role of absorption},''
  \href{http://dx.doi.org/10.1103/PhysRevD.99.064007}{{\em Phys. Rev.}
  {\bfseries D99} no.~6, (2019) 064007},
\href{http://arxiv.org/abs/1807.08840}{{\ttfamily arXiv:1807.08840 [gr-qc]}}.

\bibitem{Cardoso:2008kj}
V.~Cardoso, P.~Pani, M.~Cadoni, and M.~Cavaglia, ``{Instability of
  hyper-compact Kerr-like objects},''
  \href{http://dx.doi.org/10.1088/0264-9381/25/19/195010}{{\em Class. Quant.
  Grav.} {\bfseries 25} (2008) 195010},
\href{http://arxiv.org/abs/0808.1615}{{\ttfamily arXiv:0808.1615 [gr-qc]}}.

\bibitem{Mark:2017dnq}
Z.~Mark, A.~Zimmerman, S.~M. Du, and Y.~Chen, ``{A recipe for echoes from
  exotic compact objects},''
  \href{http://dx.doi.org/10.1103/PhysRevD.96.084002}{{\em Phys. Rev.}
  {\bfseries D96} no.~8, (2017) 084002},
\href{http://arxiv.org/abs/1706.06155}{{\ttfamily arXiv:1706.06155 [gr-qc]}}.

\bibitem{Figueras_2011}
P.~Figueras and T.~Wiseman, ``Gravity and large black holes in randall-sundrum
  ii braneworlds,''
  \href{http://dx.doi.org/10.1103/physrevlett.107.081101}{{\em Physical Review
  Letters} {\bfseries 107} no.~8, (Aug, 2011) }.
  \url{http://dx.doi.org/10.1103/PhysRevLett.107.081101}.

\bibitem{Seahra:2004fg}
S.~S. Seahra, C.~Clarkson, and R.~Maartens, ``{Detecting extra dimensions with
  gravity wave spectroscopy: the black string brane-world},''
  \href{http://dx.doi.org/10.1103/PhysRevLett.94.121302}{{\em Phys. Rev. Lett.}
  {\bfseries 94} (2005) 121302},
\href{http://arxiv.org/abs/gr-qc/0408032}{{\ttfamily arXiv:gr-qc/0408032
  [gr-qc]}}.

\bibitem{Bronnikov:2019sbx}
K.~A. Bronnikov and R.~A. Konoplya, ``{Echoes in brane worlds: ringing at a
  black hole--wormhole transition},''
\href{http://arxiv.org/abs/1912.05315}{{\ttfamily arXiv:1912.05315 [gr-qc]}}.

\end{thebibliography}\endgroup

\bibliographystyle{./utphys1}

\end{document}